\documentclass{llncs}
\usepackage{latexsym,color,graphics,times}
\usepackage{diagrams}
\usepackage{url}
\usepackage{epsfig}
\usepackage{amssymb}
\usepackage{amsmath}
\usepackage{amsfonts}

\begingroup
\catcode`\~=11
\gdef\urltilde{\lower 0.6ex\hbox{~}}
\endgroup

\newcommand{\C}{\mathcal{C}} \newcommand{\D}{\mathcal{D}}

 \newcommand{\J}{\mathcal{J}}
\newcommand{\K}{\mathcal{K}} \renewcommand{\L}{\mathcal{L}}
 
 \renewcommand{\P}{\mathcal{P}}

\title{Matching, Merging and Structural Properties of  Data Base Category}

\author{Zoran Majki\'c}
\institute{International Society for Research in Science and
Technology\\
 PO Box 2464 Tallahassee, FL 32316 - 2464 USA\\
 \email{majk.1234@yahoo.com},\\
~~~~http://zoranmajkic.webs.com/}

\newtheorem{theo}{Theorem}
\newtheorem{propo}{Proposition}
\newtheorem{coro}{Corollary}
\begin{document}


\maketitle
\begin{abstract}
Main contribution of this paper is an investigation of expressive
power of the database category $DB$. An object in this category is a
database-instance (set of n-ary relations).
 Morphisms  are not functions but have complex tree structures based on a set of complex query computations.
 They express the semantics of view-based mappings between databases.
 The  higher (logical) level scheme mappings between databases, usually written in some
 high expressive logical language, may be functorially translated into this base
 "computation" $DB$ category . The
 behavioral point of view for databases is assumed, with behavioural equivalence of databases corresponding to isomorphism of objects in $DB$ category.
  The introduced observations, which are
 view-based computations without side-effects, are based  (from Universal algebra) on monad endofunctor $T$, which is  the closure operator for
 objects and for morphisms also.
   It was shown that $DB$  is  symmetric (with a bijection between arrows and objects) 2-category, equal to its dual, complete
   and cocomplete. \\ In this paper we demonstrate that $DB$ is concrete, locally small and finitely presentable. Moreover,
   it is enriched over itself monoidal symmetric category with a tensor products for \emph{matching}, and has a parameterized \emph{merging} database
   operation. We show that it is an algebraic lattice and we define a
   database metric space and  a subobject classifier:
   thus, $DB$ category is a monoidal elementary topos.
    \end{abstract}
%
\section{Introduction}
 The relational databases are  complex structures, defined by sets of n-ary relations, and the
mappings between them are based on sets of view-mappings between the
source database $A$ to the target database $B$.  We consider the
views as an universal property for databases
  (possible observations of the information contained in some
  database).\\ We assume \emph{a
view} of a database $A$ the  relation (set of tuples) obtained by a
"Select-Project-Join + Union" (SPJRU) query $q(\textbf{x})$ where
$\textbf{x}$ is a list of attributes of this view. We denote by
$\L_A$ the set of all such queries over a database A, and by
$\L_A/_\approx$ the quotient term  algebra obtained by introducing
the equivalence relation $\approx$, such that $q(\textbf{x}) \approx
q'(\textbf{x})$ if
  both queries result with the same relation (view). Thus,
a view can be equivalently considered as a  \emph{term} of this
quotient-term algebra $\L_A/_\approx$ with carrier set of relations
in $A$ and a finite arity of their operators, whose computation
returns with a set of tuples of this view. If this query is a finite
term of this algebra it is
    called a "finitary view". Notice that a finitary view can have an infinite number of tuples
    also.\\
 Such an  \emph{instance level} database category $DB$ has been introduced first time in Technical report ~\cite{Majk03}, and used also in
 \cite{Majk03f}. General information about categories the reader can find in classic books ~\cite{McLn71}, while more information about this
 particular database category $DB$,
  with set of its objects $Ob_{DB}$ and set of its morphisms
  $Mor_{DB}$, are recently presented  in \cite{Majk04AOT}.
  In this paper we will only emphasize some of basic properties of
  this $DB$ category, in order to render more selfcontained this
  presentation.\\
Every object  (denoted by $A,B,C$,..) of this category is a
  database instance, composed by a set of n-ary relations  $a_i\in A$, $i= 1,2,...$  called also "elements of
  $A$".  Universal database instance  $\Upsilon$, is defined
  as the  union of all database instances, i.e., $\Upsilon = \{ a_i | a_i\in A, A\in Ob_{DB}\}$.
  It  is a top object of this category.\\
   It was  defined \cite{Majk04AOT} the power
  view-operator $T$, with domain and codomain equal to the set of all database
  instances, such that for any object (database) $A$, the object
  $TA$  denotes a database composed by the set of \emph{all views} of $A$. The object $TA$, for a given database instance $A$, corresponds to
   the  quotient-term  algebra $\L_A/_\approx$,
  where carrier is a set of equivalence classes of closed terms of a
well defined formulae of a relational algebra, "constructed" by
$\Sigma_R$-constructors (relational operators in SPJRU algebra:
select, project, join and union) and symbols (attributes of
relations) of a database instance $A$, and constants of
attribute-domains.
  More precisely, $TA$ is "generated" by this quotient-term algebra
  $\L_A/_\approx$, i.e., for a given evaluation of queries in $\L_A$, $Eval_A:\L_A \rightarrow TA$, which is surjective
  function, from a factorization theorem, holds that there is a
  unique bijection $is_A:\L_A/_\approx \rightarrow TA$, such that the
  following diagram in \emph{Set} category (where objects are sets, and arrows are functions) commutes
  \begin{diagram}
   \L_A  & \rTo^{Eval_{A}}   & TA\\
 \dTo^{nat_{\approx}} & \ruTo_{is_A} &\\
 \L_A/_\approx  &      &     \\
 \end{diagram}
 where the surjective function $nat_{\approx}:\L_A \rightarrow \L_A/_\approx$
 is a natural representation for the equivalence $\approx$.\\
  %
   For every object $A$ holds that $A \subseteq TA$, and $TA = TTA$, i.e., each (element) view
   of database instance $TA$ is also an element (view) of a database instance
   $A$.\\  \emph{Closed object} in $DB$ is a database $A$ such that $A = TA$. Notice that also when $A$ is  finitary
   (has a finite number of relations) but
   with at least one relation with infinite number of tuples, then $TA$ has an infinite number of relations (views of $A$),
    thus can be an infinitary object.
   It is obvious that when a domain of constants of a database is finite then both $A$ and $TA$ are finitary objects. As
   default we assume that a domain of every database is arbitrary large set but is
   finite. It is reasonable assumption for real applications.
     We have that $\Upsilon = T\Upsilon$, because every view $v\in T\Upsilon$
  is  a database instance also, thus $v\in \Upsilon$; and vice versa, every element
   $r\in \Upsilon$ is also a view of $\Upsilon$, thus $r\in T\Upsilon$.\\
      Every object (database) $A$ has also an empty relation $\bot$. The object (database) composed by only this
  empty relation is denoted by $\bot^0$ and we have that $T\bot^0
  =\bot^0= \{\bot\}$. Any empty database (a database with only empty relations) is isomorphic to this bottom object $\bot^0$.\\
   Morphisms of this category are all possible mappings
  between database instances \emph{based on views}. Elementary view-map for a given database $A$ is given by a SPCU query $f_i = q_{A_i}:A \rightarrow
  TA$.
    We will denote by $\|f_i\|$ the extension of the relation obtained by this query $q_{A_i}$. Suppose that $r_{i1},...,r_{ik} \in A$ are
the relations used for computation of this query, and that  the
corespondent algebraic term $~\widehat{q_i}$ is a function (it is
not a T-coalgebra) $\widehat{q_i}:A^k \rightarrow TA$, where $A^k$
is k-th cartesian product of $A$. Then, $\|q_{A_i}\| =
\widehat{q_i}(r_{i1},...,r_{ik})$. Differently from this algebra
term  $~\widehat{q_i}$ which is a function, a view-map $q_{A_i}:A
\rightarrow  TA$, which is \emph{a T-coalgebra}, \emph{is not} a function.\\
Consequently, an atomic morphism $f:A \rightarrow B$, from a
database $A$ to
database $B$, is a set of such view-mappings, thus \emph{it is not} generally a function.\\
We can introduce two functions, $\partial_0, \partial_1:Mor_{DB}
\rightarrow \P(\Upsilon)$ (which are different from standard
category functions $dom, cod:Mor_{DB} \rightarrow Ob_{DB}$), such
that for any view-map $q_{A_i}:A \longrightarrow TA$, we have that
   $\partial_0(q_{A_i}) = \{ r_1,..., r_k\} \subseteq A$  is a subset of relations of $A$
    used as arguments by this query $q_{A_i}$ and
   $\partial_1(q_{A_i})= \{v\}, v \in TA$ ($v$ is a resulting view of a query
   $q_{A_i}$). In fact, we have that they are functions $~\partial_0,
\partial_1:Mor_{DB} \rightarrow \P(\Upsilon)$ (where $\P$ is a powerset
operation), such that for any morphism $f:A \rightarrow B$ between
databases $A$ and $B$, which is a set of view-mappings $q_{A_i}$
such that $\|q_{A_i}\| \in B$, we have that $\partial_0(f) \subseteq
A$ and $\partial_1(f) \subseteq TA \bigcap B \subseteq B$.
  Thus,  we have
   \[
\partial_0(f)=  \bigcup_{ q_{A_i}\in f}
\partial_0 (q_{A_i}) \subseteq dom(f) = A,~~
~~ \partial_1 (f) = \bigcup_{ q_{A_i}\in f}
\partial_1 (q_{A_i})\subseteq cod(f) = B~~
\]
We may define equivalent (categorically isomorphic) objects
(database instances) from the \emph{behavioral point of view based
on observations}:  each arrow (morphism) is composed by a number of
"queries" (view-maps), and each query may be seen as an
\emph{observation} over some database instance (object of $DB$).
Thus, we can characterize each object in $DB$ (a database instance)
by its behavior according to a given set of observations.  Thus
databases $A$ and $B$ are equivalent (bisimilar) if they have the
same set of its observable internal states, i.e. when $TA$ is equal
to $TB$: $~~~~
A\approx B~~iff~~TA = TB $.\\
Basic properties of this database category $DB$ as its symmetry
(bijective correspondence between arrows and objects, duality ($DB$
is equal to its dual $DB^{OP}$) so that each limit is also colimit
(ex. product is also coproduct, pullback is also pushout, $\bot^0$
is zero objet, that is, both initial and terminal object, etc..),
and that it is a 2-category has been demonstrated in
\cite{Majk03,Majk04AOT}.\\
Generally, database mappings are not simply programs from values
(relations) into computations (views) but an equivalence of
computations: because of that each mapping, from any two databases A
and B, is symmetric and gives a duality property to the category
$DB$. The denotational semantics of database mappings is given by
morphisms of the Kleisli category $DB_T$ which may be "internalized"
in $DB$ category as "computations" \cite{MaBh10}.\\
 The product $A\times B$ of a databases
$A$ and $B$ is equal to their coproduct $A+B$, and the semantics for
them is that we are not able to define a view by using relations of
both databases, that is, these two databases have independent DBMS
for query evaluation. For example, the creation of exact copy of a
database $A$ in another DB server corresponds to the database $A +
A$. In this paper we will introduce the denotational semantics for
other two fundamental database operations as matching and merging
(and data federation), and we will show other more
advanced properties.\\
Plan of this paper is the following: After brief technical
preliminaries taken from \cite{Majk03,Majk04AOT,Majk08db}, in
Section 2 we will consider some Universal algebra considerations and
relationships of $DB$ and standard $Set$ category. In Section 3 we
will introduce the categorial (functors) semantics for two basic
database operations: matching and merging, while in Section 4 we
will define the algebraic database lattice and will show that $DB$
is concrete, small and locally finitely presentable (lfp) category.
In Section 5 we will show that $DB$ is also V-category enriched over
itself. Finally in Section 6 we will develop a metric space and a
subobject classifier for this category, and we will show that it is
a weak monoidal topos.
\subsection{Technical preliminaries}
Based on atomic morphisms (sets of view-mappings)
\cite{Majk04AOT,Majk08db} which are complete arrows (c-arrows), we
obtain that their composition generates tree-structures, which can
be incomplete (p-arrows), in the way that for a composed arrow $h =
g\circ f:A \rightarrow C$, of two atomic arrows $f:A \rightarrow B$
and $g:B\rightarrow C$, we can have the situations where
$\partial_0(f) \subset
\partial_0(h)$.
\begin{definition}\label{def:morphisms}  The following BNF defines the set $Mor_{DB}$ of all morphisms in
DB:\\
  $~~~p-arrow \textbf~{:=} ~c-arrow ~|~c-arrow \circ c-arrow~$ (for any two c-arrows  $f:A\longrightarrow B$ and $g:B\longrightarrow C~$)\\
 $~~~morphism \textbf~{:=} ~p-arrow ~|~c-arrow \circ p-arrow~$ (for any p-arrow $f:A\longrightarrow B$ and c-arrow $g:B\longrightarrow C$)\\
\\whereby the composition of two arrows, f (partial) and g (complete),
we obtain the following p-arrow (partial arrow) $h = g \circ
f:A\longrightarrow C$
\[
\ h = g\circ f = \bigcup_{ q_{B_j}\in ~g ~\& ~\partial_0 (q_{B_j})
\bigcap
\partial_1 (f) \neq \emptyset } \{q_{B_j}\}~~~~\circ
\]
\[
\circ ~~~~ \bigcup_{q_{A_i}\in ~f ~\&~\partial_1 (q_{A_i})=
\{v\}~\&~ v \in ~\partial_0 (q_{B_j}) } \{q_{A_i}(tree)\}~~
\]
$= \{q_{B_j} \circ \{q_{A_i}(tree)~|~\partial_1(q_{A_i}) \subseteq
\partial_0(q_{B_j})\}~|~q_{B_j}\in ~g ~\&
~\partial_0 (q_{B_j}) \bigcap
\partial_1 (f) \neq \emptyset \}$\\
$= \{q_{B_j}(tree)~|~q_{B_j}\in ~g ~\& ~\partial_0 (q_{B_j}) \bigcap
\partial_1 (f) \neq \emptyset\}$
\\where   $q_{A_i}(tree)$ is the  tree of the morphisms f below
$q_{A_i}$.\\
We define the semantics of mappings by function
 $B_T:Mor_{DB}\longrightarrow Ob_{DB}$, which, given any mapping
 morphism
 $f:A\longrightarrow B~$ returns with the set of views ("information flux") which are
 really "transmitted" from the source to the target object.\\
 1. for atomic morphism, $ \widetilde{f} = B_T(f)~\triangleq
 T\{\|f_i\|~|~f_i \in f \}$.\\
 2. Let $g:A \rightarrow B$ be a morphism with a flux
 $\widetilde{g}$, and $f:B \rightarrow C$ an atomic morphism with
 flux $\widetilde{f}$ defined in point 1, then  $~~\widetilde{f \circ g}
 = B_T(f \circ g) ~\triangleq \widetilde{f}\bigcap
 \widetilde{g}$.\\
 We introduce an equivalence relation over
morphisms by, $ ~~~~f \approx g ~~~~iff~~~~ \widetilde{f} =
\widetilde{g}$.
 \end{definition}
Notice that between any two databases $A$ and $B$ there is at least
an "empty" arrow $f:A \rightarrow B$ such that $\partial_0(f) =
\partial_1(f) = \widetilde{f} = \bot^0$.
 Thus we have the following fundamental properties:
\begin{propo}\label{prop-morphism} Any mapping  morphism
 $f:A\longrightarrow B~$  is a closed object in DB, i.e., $~~\widetilde{f} =
T\widetilde{f}$, such that $\widetilde{f} \subseteq TA \bigcap TB$, and\\
  1. each arrow  such that
$~\widetilde{f} =   TB$ is an epimorphism $f:A\twoheadrightarrow B$,\\
  2. each arrow  such that $~\widetilde{f}=
  TA$ is a monomorphism $f:A\hookrightarrow B$,\\
  3. each monic and epic arrow is an isomorphism.
\end{propo}
If $f$ is epic then $TA \supseteq TB$;  if it is monic then $TA
\subseteq TB$. Thus we have an isomorphism of two objects
(databases), $A \simeq B$, iff $TA = TB$, i.e., when they are
observationally equivalent $A\approx B$.
Thus, for any database $A$ we have that $A\simeq TA$.\\
 Let us extend the notion of the type operator $T$ into a
notion of the power-view endofunctor in $DB$ category:
\begin{theo} \label{th:endofunctor}  There exists an endofunctor $T =
(T^0,T^1):DB \longrightarrow DB$, such that
\begin{enumerate}
  \item for any object A, the object component $T^0$ is
  equal to the type operator T, i.e.,  $~~~~T^0(A)\triangleq TA$
  \item for any morphism $~f:A\longrightarrow B$, the arrow
  component $T^1$ is  defined by
  \[
  \ T(f) \triangleq T^1(f) =  \bigcup_{ \partial_0(q_{TA_i})
  =\partial_1(q_{TA_i}) =\{v\} ~\& ~v \in ~ \widetilde{f}} \{q_{TA_i}:TA
  \rightarrow TB\}
  \]
  \item Endofunctor T preserves the properties of arrows, i.e., if a
  morphism $f$ has a property P (monic, epic, isomorphic), then
  also $T(f)$ has the same property: let $P_{mono}, P_{epi}
  ~and~\\P_{iso}$ are monomorphic, epimorphic and isomorphic
  properties respectively, then the following formula is true\\
  $\forall(f\in Mor_{DB})(P_{mono}(f)\equiv P_{mono}(Tf)$ and $P_{epi}(f)\equiv
  P_{epi}(Tf)$  and  $P_{iso}(f)\equiv P_{iso}(Tf)$.
  \end{enumerate}
\end{theo}
\textbf{Proof:} It is easy to verify that $T$ is a 2-endofunctor and
to see that $T$ preserves properties of arrows: for example, if
$P_{mono}(f)$ is true for an arrow $f:A\longrightarrow B$, then
$\widetilde{f} = TA$ and $\widetilde{Tf}= T \widetilde{f} = T(TA) =
TA$, thus $P_{mono}(Tf)$ is true. Viceversa, if $P_{mono}(Tf)$ is
true then $\widetilde{Tf} =T \widetilde{f} = T(TA)$, i.e.,
$\widetilde{f} = TA$ and, consequently, $P_{mono}(f)$ is true.
\\$\square$\\
 The equivalence
relations on objects and morphisms are based on the "inclusion"
Partial Order (PO) relations, which define the DB as a 2-category:
\begin{propo}\label{prop:POCat}  The subcategory $DB_I \subseteq DB$ , with $Ob_{DB_I}
= Ob_{DB}$ and with only monomorphic arrows, is a Partial Order
category with PO relation of "inclusion" $A \preceq B$ defined by a
monomorphism $f:A\hookrightarrow B$.  The "inclusion" PO relations
for objects and arrows are defined as follows:
\[
 A \preceq B ~~~~ iff ~~~~ TA \subseteq TB
\]
\[
 f \preceq g ~~ ~~iff ~~~~ \widetilde{f } \preceq
 \widetilde{g}~~~~(i.e.,\widetilde{f } \subseteq
 \widetilde{g}~~)
\]
      they determine observation equivalences, i.e.,
\[
 A \backsimeq B ~~(i.e.,~~ A \thickapprox B)~~~~~~ iff
 ~~ ~~A\preceq B~~and ~~B \preceq A
\]
\[
 f \thickapprox g ~~~~ iff ~~~~ f  \preceq g ~~and ~~ g \preceq f
 ~
\]
The power-view endofunctor $T:DB\longrightarrow DB$ is a
2-endofunctor and the closure operator for this PO relation: any
object A such that $A = TA$ will be called "closed object".\\
 $DB$ is a 2-category, 1-cells are its ordinary morphisms, while
2-cells (denoted by $\sqrt{\_}~$) are the arrows between ordinary
morphisms : for any two morphisms $~f,g:A\longrightarrow B$ , such
that $f \preceq g~$, a 2-cell arrow is the "inclusion"
$\sqrt{\alpha}:f\frac{\preceq}{\longrightarrow}g $. Such a 2-cell
arrow is represented by an ordinary monic arrow in DB,
$~~\alpha:\widetilde{f} \hookrightarrow \widetilde{g}$.
\end{propo}
The following duality theorem tells that, for any commutative
diagram in $DB$ there is also the same commutative diagram composed
by the equal objects and inverted equivalent arrows: This
"bidirectional" mappings property of $DB$ is a consequence of the
fact that the composition of arrows is semantically based on the
set-intersection commutativity property for "information fluxes" of
its arrows. Thus \emph{any limit diagram} in $DB$ has also its
\emph{"reversed" equivalent colimit diagram} with equal objects,
\emph{any universal property }has also its \emph{equivalent
couniversal property }in $DB$.
\begin{theo} \label{th:duality} there exists the controvariant functor
$\underline{S}= (\underline{S}^0,\underline{S}^1):DB \longrightarrow
DB$ such that
\begin{enumerate}
  \item $\underline{S}^0$ is the identity function on objects.
  \item for any arrow in DB, $f:A\longrightarrow B$ we have $\underline{S}^1(f):B\longrightarrow
  A$, such that $\underline{S}^1(f) \triangleq f^{inv}$, where $f^{inv}$
  is (equivalent) reversed morphism of $~f~~~$ (i.e., $\widetilde{f^{inv}}= \widetilde{f}$),\\
  $f^{inv}= is^{-1}_A \circ(Tf)^{inv}\circ is_B~$  with
\[
  \ (Tf)^{inv} ~\triangleq \bigcup_{ \partial_0(q_{TB_j})
   = \partial_1(q_{TB_j}) =\{v\} ~\& ~v \in~  \widetilde{f}} \{q_{TB_j}:TB
   \rightarrow TA \}
\]
  \item The category DB is equal to its dual category $DB^{OP}$.
\end{enumerate}
\end{theo}
\textbf{Proof:} it can be found in \cite{Majk08db}
\\$\square$\\
\section{Universal algebra considerations \label{section:universal}}
In order to explore universal algebra properties for the category
 $DB$ \cite{Majk09f}, where, generally, morphisms are not functions (this fact complicates a definition of mappings from its morphisms
 into homomorphisms of the category of $\Sigma_R$-algebras), we
 will use an equivalent to $DB$ "functional" category, denoted by
 $DB_{sk}$, such that  its arrows can be seen as total
 functions.
 \begin{propo}\label{prop-skeletal}
 Let us denote by $DB_{sk}$ the full skeletal
 subcategory of DB, composed by  closed objects only.\\
 Such a category is equivalent to the category DB, i.e., there
 exists an adjunction of a surjective functor $T_{sk}:DB\longrightarrow DB_{sk}$ and an
 inclusion functor $In_{sk}:DB_{sk}\longrightarrow DB$ such that
 $T_{sk}In_{sk} = Id_{DB_{sk}}$ and $In_{sk}T_{sk}\simeq
 Id_{DB}$.\\
  There exists the faithful forgetful functor
 $F_{sk}:DB_{sk}\longrightarrow Set $, and $F_{DB}= F_{sk}\circ T:DB\longrightarrow
 Set$,  thus $~DB_{sk}~$ and $DB$ are  \textsl{concrete} categories.
  \end{propo}
\textbf{Proof:}   Let us define $T^0_{sk} = T^0$ and $T^1_{sk} =
T^1$,   while $In^0_{sk}$ and $In^1_{sk}$ are two identity
functions.
  It is easy to verify that these two categories are equivalent. In fact,
  there exists an adjunction $<T_{sk}, In_{sk}, \eta_{sk}, \varepsilon_{sk}>:DB\longrightarrow DB_{sk}$, because of the bijection
  $DB_{sk}(T_{sk}A,B)\simeq DB(A, In_{sk}B)$ which is natural in
  $A \in DB$ and $B\in DB_{sk}$ ($B$ is closed, i.e., $B = TB$). In
  facts,
$DB_{sk}(T_{sk}A,B) = \{ \widetilde{f} ~|~f:TA \longrightarrow B\} =
\{ \widetilde{g} ~|~g:A \longrightarrow B\} = \{ \widetilde{g}
~|~g:A \longrightarrow In_{sk}B\} =DB(A, In_{sk}B)$.
   The skeletal category $DB_{sk}$ has closed
  objects only, so, for any two closed objects $TA$ and $TB$, each arrow
  between them  $f:TA\longrightarrow TB$ can be expressed in a
  following  "total" form $~f_T = f~$ (such that $\partial_0(f_T) = TA$)
  \[
  \ f_T \triangleq  \bigcup_{ \partial_0(q_{TA_i})
  =\partial_1(q_{TA_i}) = \{v\}~\&~ v \in ~ \widetilde{f}}
  \{q_{TA_i}\}
    \bigcup_{ \partial_0(q_{TA_i}) = \{v\}~\&~ v \notin \widetilde{f}~\&~\partial_1(q_{TA_i}) = \perp^0 }
    \{q_{TA_i}\}
  \]
 Thus, a morphism $f_T$ can be seen as a (total) function from $TA$ to $TB$, such
 that for any $v\in TA$ we have that $f_T(v) = v$ if $v\in
 \widetilde{f}$, $\perp$ otherwise. Such an analog property is valid
  for its reversed equivalent morphism $f^{inv}_T:TB \longrightarrow
 TA$ also.\\
 Let us define the functor $F_{sk}:DB_{sk}\longrightarrow Set $ by:
 $F^0_{sk}$ an identity function on objects and for any arrow
 $f:TA \longrightarrow TB$ in $DB_{sk}$ we obtain a function $g = F^1_{sk}(f)$ from a set $TA$ to a set $TB$ such
 that for any relation $v \in TA$,
 $g(v) \triangleq \{v, $if $v\in \widetilde{f}; \bot$ otherwise$ \}$. It is easy to verify that $F^1_{sk}(f) = F^1_{sk}(h)$
 implies $f = h$, i.e., $F_{sk}$ is a faithful functor and also $F_{DB} = F_{sk}\circ
 T$ is a faithful. Thus $DB_{sk}$ and $DB$ are concrete categories.
\\$\square$\\
 In a given inductive definition one defines a value of a function
 (in our example the endofunctor $T$) on all (algebraic)
 constructors (relational operators). What follows is based on the
 fundamental results of the Universal algebra ~\cite{Cohn65}.\\
 Let $\Sigma_R$ be a finitary signature (in the usual algebraic
 sense : a collection $F_\Sigma $ of function \emph{symbols} together with
 a function $~ar:F_\Sigma \longrightarrow N$ giving the finite arity
 of each function symbol) for a single-sorted (sort of relations) relational
 algebra.\\
 We can speak of $\Sigma_R$-equations and their satisfaction in a
$\Sigma_R$-algebra, obtaining the notion of a $(\Sigma_R,
E)$-algebra theory. In a special case, when $E$ is empty, we obtain
a purely syntax version of Universal algebra, where $\K$ is a
category of all $\Sigma_R$-algebras, and the
quotient-term algebras are simply term algebras. \\
An \emph{algebra} for the algebraic theory (type) $(\Sigma_R, E )$
is given by a set $X$, called the \emph{carrier} of the algebra,
together with interpretations for each of the function symbols in
$\Sigma_R$. A function symbol $f \in \Sigma_R$ of arity $k$ must be
interpreted by a function $\widehat{f}_X:X^k \longrightarrow X$.
Given this, a term containing $n$ distinct variables gives rise to a
function $X^n\longrightarrow X $ defined by induction on the
structure of the term. An algebra must also satisfy the equations
given in $E$ in the sense that equal \emph{terms} give rise to
identical functions (with obvious adjustments where the equated
terms do not contain exactly the same variables).
 A \emph{homomorphism} of algebras from an algebra X to an algebra Y is
given by a function $ g:X\longrightarrow Y $ which commutes with
operations of the algebra $g(\widehat{f}_X(x_1,..,x_k)) =
~\widehat{f}_Y(g(x_1),..,g(x_k))$.\\
 This generates a variety category $\K$ of all relational algebras.
Consequently, there is a bifunctor $E:DB^{OP}_{sk}\times \K
\longrightarrow Set$ (where $Set$ is the category of sets), such
that for any database instance $A$ in $DB_{sk}$ there exists the
functor $~E(A,\_ ~):\K\longrightarrow  Set$ with an universal
element $(U(A), \varrho)$, where  $\varrho\in E(A,U(A))$ ,
 $\varrho:A \longrightarrow U(A)$ is an inclusion function and
 $U(A)$ is a free algebra over $A$ (quotient-term algebra generated by a carrier
 database instance $A$), such that for any function $f\in E(A,X)$
 there is a unique homomorphism $~h~$ from the free algebra $U(A)$ into an
 algebra $X$, with $~f = E(A,h)\circ \varrho$.\\
 From the so called "parameter theorem" we obtain that there exists:
\begin{itemize}
  \item  a unique universal functor $~U:DB_{sk} \longrightarrow \K~$ such
 that for any given database instance $A$ in $DB_{sk}$ it returns with the free
$\Sigma_R$-algebra  $U(A)$ (which is a quotient-term algebra, where
a carrier is a set of equivalence classes of closed terms of a well
defined formulae of a relational algebra, "constructed" by
$\Sigma_R$-constructors (relational operators: select, project, join
and union SPJRU) and symbols (attributes and relations) of a
database instance $A$, and constants of attribute-domains. An
alternative for $U(A)$ is given by considering  $A$ as a set of
variables rather than a set of constants, then we can consider
$U(A)$ as being a set of \emph{derived operations} of arity $A$ for
this theory. In either case the operations are interpreted
syntactically $\widehat{f}([t_1],...,[t_k]) = [f(t_1,...,t_k)]$,
where, as usual, brackets denote equivalence classes), while, for
any "functional" morphism (correspondent to the total function
$F^1_{sk}(f_T)$ in Set, $F_{sk}:DB_{sk} \longrightarrow Set$)
$~f_T:A\longrightarrow B$ in $DB_{sk}$ we obtain the homomorphism $
f_H = U^1(f_T)$ from the $\Sigma_R$-algebra $U(A)$ into the
$\Sigma_R$-algebra $U(B)$, such that for any term $\rho (a_1,..,a_n)
\in U(A)$, $\rho \in \Sigma_R$, we obtain $f_H(\rho (a_1,..,a_n)) =
\rho (f_H(a_1),...,f_H(a_n))$, so, $f_H$ is an identity function for
algebraic operators and it is equal to the function $F^1_{sk}(f_T)$
for constants.
  \item its adjoint forgetful functor $F:\K \longrightarrow DB_{sk}$,
  such that for any free algebra $U(A)$ in $\K$ the object $F \circ U(A)$
  in $DB_{sk}$ is equal to its carrier-set $A$ (each term $\rho (a_1,...,a_n) \in U(A)$
   is evaluated into a view of this closed object $A$ in $DB_{sk}$)
  and for each arrow $U^1(f_T)$ holds
  that $F^1U^1(f_T) = f_T$, i.e., we have that $FU = Id_{DB_{sk}}$ and $UF =
  Id_{\K}$.
  \end{itemize}
 Consequently,  $U(A)$ is a quotient-term algebra,
where carrier is a set of equivalence classes of closed terms of a
well defined formulae of a relational algebra, "constructed" by
$\Sigma_R$-constructors (relational operators in SPJRU algebra:
select, project, join and union) and symbols (attributes of
relations) of a
database instance $A$, and constants of attribute-domains.\\
It is immediate from the universal property that the map $A \mapsto
U(A)$ extends to the endofunctor $F \circ U:DB_{sk}\longrightarrow
DB_{sk}$. This functor carries \emph{monad structure} $(F \circ U,
\eta, \mu)$ with $F \circ U$ an equivalent version of $T$ but for
this
skeletal database category $DB_{sk}$.\\
 The natural
  transformation $\eta $ is given by the obvious "inclusion" of
  $A$ into $F\circ U(A): a\longrightarrow [a]$ (each view $a$ in an closed object $A$
   is an equivalence class of all algebra terms which produce this view).
    Notice that the natural transformation $\eta$ is the unit of this adjunction of $U$ and $F$,
    and that it corresponds to an inclusion function in $Set$, $\varrho:A \longrightarrow U(A)$, given above. The
  interpretation of $ \mu $ is almost equally simple. An element
  of $(F\circ U)^2(A) $ is an equivalence class of terms built up
  from elements of $F\circ U(A)$, so that instead of $t(x_1,...,x_k)$, a typical element of $(F\circ U)^2(A) $ is
  given by the equivalence class of a term $ t([ t_1 ],...,[ t_k ])$.
  The transformation $\mu $ is defined by map $ [t([ t_1 ],...,[ t_k ])]~\mapsto~ [t(t_1 ,.., t_k )]$. This make sense because a
  substitution of provably equal expressions into the same term
  results in provably equal terms.
\section{Matching and Merging database operations}
In this section we will investigate the properties of  $DB$ category
and, especially, its functorial constructs for the algebraic
high-level operators over databases: for example \cite{Majk09a},
matching, merging,etc..
%
\subsection{Matching tensor product}
 Since the data residing in different databases may have inter-dependencies (they
are based on the partial overlapping between  databases, which is
information about a common part of the world) we can define such an
(partial) overlapping by morphisms of the category $DB$:
"information flux" of each mapping between two objects $A$ and $B$
in $DB$ is just a subset of this overlapping  between these two
databases, denoted by $A \otimes B$. It is "bidirectional" ,i.e.,(by
duality) for any mapping $f~$ from $A$ into $B$ there exists an
\emph{equivalent} mapping $f^{inv}~$ from $B$ into $A$. This
overlapping represents the common matching between these two
databases, and is equal to the maximal "information flux" which can
be defined between these two databases. Consequently,  we can
introduce formally a denotational semantics for database matching
operation $\otimes$, as follows:
\begin{propo}\label{prop-matching}
DB is a strictly symmetric idempotent monoidal category $(DB,
\otimes, \Upsilon, \alpha,\\ \beta, \gamma )~$, where $~\Upsilon~$
is the total object for a given universe for databases, with the
"matching" tensor product $ \otimes :DB\times DB\longrightarrow DB$
defined as follows:
\begin{enumerate}
  \item for any two database instances (objects) A and B, $A\otimes B$ is the
  overlapping (matching) between A and B, defined by the
  bisimulation equivalence relation (i.e., by their common
  observations): $~~~~A\otimes B \equiv \otimes (A,B) \triangleq
  (\bigcap \cdot T)(A,B) =  TA \bigcap TB$
  \item for any two arrows $f:A\rightarrow C~$ and $g:B\rightarrow
  D$,
  \[
  \ f\otimes g \equiv \otimes(f,g) ~\triangleq \bigcup_{ \partial_0(q_{(A\otimes B)_i})
   =\partial_1(q_{(A\otimes B)_i}) = \{v\}~ \& ~v \in~
  \widetilde{f} \bigcap \widetilde{g}} \{q_{(A\otimes B)_i}\}~
  \]
  \item for any two objects A,B, every morphism $f:A\rightarrow
  B~$ satisfy $~\perp^0\subseteq \widetilde{f} \subseteq A\otimes B$.
\end{enumerate}
\end{propo}
\textbf{Proof:} It is easy to verify that $~\otimes~$ is monoidal
bifunctor , with
natural isomorphic transformations (which generate an identity arrow for each object in $DB$):\\
$~~~\alpha:(\_ \otimes \_ )\otimes \_ \longrightarrow \_\otimes (\_ \otimes \_ )$, $~~~~~~$ associativity\\
 $~~~\beta:\Upsilon \otimes \_ \longrightarrow I_{DB}$, $~~~~~~$left identity\\
 $~~~\gamma: \_ \otimes \Upsilon  \longrightarrow I_{DB}$, $~~~~~~$ right identity\\
 such that $A\otimes B = B\otimes A$, $~A\otimes
\Upsilon = \Upsilon\otimes A = A$, $~A\otimes A \simeq A$. For any
morphism $f:A\longrightarrow B$, from $~\widetilde{f} \subseteq TA$
and $~\widetilde{f} \subseteq TB$ we obtain $~\widetilde{f}
\subseteq  TA\bigcap TB = A\otimes B$.\\
Moreover, for any database $A$ we have that $A \otimes \bot^0 =
\bot^0$.
\\$\square$\\
Tensor product $\otimes$ of the monoidal category $DB$ \emph{is not}
unique in contrast with the Cartesian product (we can have $A\otimes
B = C\otimes B$ such that $ C = A\bigcup A_1 \succ A \succeq B$).\\
Notice that each $~A\otimes B~$ is a closed object (intersection of
two closed objects $TA$ and $TB$), and that the "information flux"
of any morphism from $A$ to $B$ is a closed object included in this
"maximal information flux" (i.e., overlapping) between $A$ and $B$.
Two completely disjoint databases have as overlapping (the maximal
possible "information interchange flux") the empty bottom object
$\perp^0$.
\begin{propo}\label{prop-monoid} Each object A together with two arrows,
an isomorphism $\mu_A:A\otimes A \longrightarrow A$ and an
epimorphism $\eta_A:\Upsilon \twoheadrightarrow A$ , is a
\textsl{monoid} in the monoidal category $(DB, \otimes, \Upsilon,
\alpha, \beta, \gamma)$.
\end{propo}
\textbf{Proof:} It is easy to verify that is valid $\mu_A \circ
(\mu_A \otimes id_A)\circ \alpha_{A,A,A} = \mu_A \circ (id_A \otimes
\mu_A )$ and $ \beta_A = \mu_A \circ (\eta_A \otimes id_A)$,
$\gamma_A = \mu_A \circ (id_A \otimes \eta_A)$.
\begin{propo}\label{prop-matching} The following properties for arrows in DB are valid:
\begin{itemize}
  \item for any two objects A, B, the arrow $h:A\longrightarrow B$ such that
  $\widetilde{h} = A\otimes B$ is a \textsl{principal} morphism.
  \item for any monomorphism $f:A \hookrightarrow B$ and its
  reversed epimorphism $f^{inv}:B\twoheadrightarrow A$, $(f, f^{inv})$ is a \textsl{retraction} pair.
  \item for any object A there is a category of \textsl{idempotents} on A (denoted by $Ret_A$) defined as follows:\\
  1. objects of $Ret_A$ is the set of all arrows from A into A,
  i.e., $Ob_{Ret_A} = DB(A,A)$.\\
  2. for any two objects $f,g \in Ob_{Ret_A}$ arrows between them
  are defined by the bijection $~~\varphi:DB(\widetilde{f}, \widetilde{g})
  \simeq Ret_A(f, g)~~$ such that for any $h\in DB(\widetilde{f},
  \widetilde{g})$ holds $ h \approx \varphi(h)$.
\end{itemize}
\end{propo}
\textbf{Proof:} 1.  for any $h:A\longrightarrow B$ with
$\widetilde{h} = A\otimes B$ holds that $\forall f:A\longrightarrow
B, \exists g:A\longrightarrow A $, such that $f = h\circ g$ (in fact
for $g \approx f$ it is satisfied).\\
2. for any monomorphism $f:A \hookrightarrow B$  and epimorphism
$f^{inv}:B\twoheadrightarrow A$ ($f^{inv} \approx f$) holds that
$f^{inv}\circ f = id_A$ (in fact, $\widetilde{f^{inv}\circ f} =
\widetilde{f^{inv}}\bigcap \widetilde{f} = \widetilde{f} = TA =
\widetilde{id_A}$).\\
3. for each $f:A \hookrightarrow A$ holds $f\circ f = f$. Thus, it
is idempotent and, consequently, an object in $Ret_A$. For any $h\in
DB(\widetilde{f},  \widetilde{g})$ in $DB$,  the arrow $k =
\varphi(h)\in Ret_A(f, g)$, such that $h\approx k$ , satisfies $k =
g\circ k\circ f$ in $DB$. Demonstration: from
$h:\widetilde{f}\longrightarrow \widetilde{g}$ it holds that
$\widetilde{h}\subseteq T\widetilde{f}\bigcap T\widetilde{g} =
\widetilde{f} \bigcap \widetilde{g}$, consequently $\widetilde{k} =
\widetilde{k} \bigcap \widetilde{f} \bigcap \widetilde{g}=
\widetilde{g\circ k\circ f}$.\\
Notice that for any 2-cell $h:\widetilde{f}\preceq \widetilde{g}$ we
have that $\varphi(h)= f \in Ret_A(f, g)$ (in fact, $h$ is
monomorphism, thus, $\widetilde{h} = \widetilde{f}$ and also
$\widetilde{h} = \widetilde{\varphi(h)}$, thus
$\widetilde{\varphi(h)} = \widetilde{f}$, i.e., $\varphi(h)= f$).
\\$\square$
\subsection{Merging operator}
Merging of two databases $A$ and $B$ is similar to the concept of
union of two databases in one single database. As we will show, this
similarity  corresponds to an isomorphism in $DB$. That is, the
union of two databases is isomorphic to the database obtained by
their merging, from the behavioral point of view. Any view which can
be obtained from union of two databases, can also be obtained from
merging these two databases, and vice versa.\\
In what follows, similarly to matching tensor products which, for
any two given databases, returns with only closed objects, also the
merging operator will return with closed objects. As we will see
these two operators will result as meet and joint operators of
complete algebraic database lattice where $\bot^0$ and $\Upsilon$
are bottom and top elements respectively.
\begin{propo}\label{prop-merging}
For any fixed database (object) $A$ in $DB$ we define the
parameterized  "merging with A" operator as an endofunctor $ A \oplus
\_ :DB \longrightarrow DB$  as follows:
\begin{enumerate}
  \item for any database instance (object)  $B$, $A\oplus B$ is a merging of A and B, defined by the
  bisimulation equivalence relation: $~~~~A\oplus B \equiv \oplus (A,B)
  \triangleq (T \cdot \bigcup)(A,B) =  T(A \bigcup B)$
  \item for any  arrow  $f:B\rightarrow
  C$, $~~~~ A \oplus (f) \triangleq (id_A \bigcup f): A\oplus B \rightarrow A\oplus
  C$,\\
  such that $~ \widetilde{A \oplus (f)} = A \oplus \widetilde{f}$.
\end{enumerate}
\end{propo}
\textbf{Proof:} It is easy to verify that
$T(A \bigcup B) = T(TA \bigcup TB)$, that is $A \oplus B = TA \oplus
TB$, and $A \bigcup B \simeq TA \bigcup TB \simeq A \oplus B$.
\\Now we can verify that $~A \oplus \_~$ is an endofunctor. In fact,
for any identity arrow $id_B:B \rightarrow B$, we have that $ A
\oplus (id_B) = id_A \bigcup id_B$, so that $\widetilde{A \oplus
(id_B)} = T(\widetilde{id_A} \bigcup \widetilde{id_B}) = T(TA
\bigcup TB) = T(A \bigcup B) = \widetilde{id_{A\oplus B}}$.
 Consequently, for identity arrows holds
functorial property, $ A \oplus (id_B) = id_{A\oplus B}$.\\
From the fact that for any object (database) $B$, we have that $A
\subseteq A \oplus B$, each arrow resulting by application of this
endofunctor contains a sub arrow $id_A$. Thus, given two arrows
$f:B\rightarrow C$ and $g:C\rightarrow D$, we have the compositional
endofunctors property, $A \oplus (g)\circ A \oplus (f) = (id_A
\bigcup g) \circ (id_A \bigcup f) = id_A \bigcup (g \circ f) = A
\oplus (g \circ f)$.\\
 Moreover,  we have that  $A\oplus B = B\oplus A$, $A \oplus \Upsilon
= \Upsilon$, $~A\oplus \bot^0 =  A$,  $~A\oplus TA = TA$ and
$~A\oplus A \simeq A$.
\\$\square$\\
Matching and merging operators are dual operators in the category
$DB$: in fact they are also dual lattice operators (meet and join
respectively) w.r.t. the database ordering $\preceq$,
as we will show in what follows.\\
Notice that for the objects in database category, the commutative
operation of merging $\oplus = T \cdot \bigcup$ is a generalization
of the set union operation $\bigcup$ in the category
of sets $Set$.\\
 \textbf{Remark:} \emph{Data federation} of two
databases $A$ and $B$ is their union, that is a database $A \bigcup
B$. It is easy to see that $A \bigcup B \simeq A \oplus B$, that is,
from the behavioral point of view, data federation is equivalent to
data merging, that is for any query over data federation $A \bigcup
B$, which returns with a view $r$, there exists a query over data
merging $A \oplus B$ which returns with the same view $r$; and vice
versa.
\section{Algebraic database lattice}

 We have seen that the set of all closed objects (i.e., objects of the skeletal
 category $DB_{sk}$, equivalent to $DB$), denoted by $\C
 \triangleq Ob_{DB_{sk}}$, defines a \emph{closed set system} \\$<\Upsilon, \C>$, where $\Upsilon$ is a closed "total" object
 (a merging, or up to isomorphism a union (we have that $A \bigcup B \simeq A \oplus B$), of all objects
 (database instances) of $DB$), correspondent to the closure operator $T$. Thus, the
 lattice $<\C, \subseteq>$ with respect to the set-inclusion $ \subseteq
 $ is a complete lattice ~\cite{Cohn65}. We recall the fact that a complete lattice is a poset $P$ such that for any subset $S$
  both $infS$ (greatest lower bound) and $supS$ (least upper bound) exist in $P$:
  for any $A, B, C \in \C$, and binary operations "join" $\bigvee$ and "meet"
  $\bigwedge$  (in the case of $Set$ category these operators are set-union and set-intersection respectively, while for $DB$ category
  we will show that they are merging and matching operators respectively), the following identities are satisfied \\
 1)$~~A \bigvee B = B\bigvee A, ~~A\bigwedge B = B\bigwedge A$
  commutative laws\\
 2)$~~A\bigvee (B\bigvee C) = (A\bigvee B)\bigvee
  A, ~~A\bigwedge(B\bigwedge C) = (A\bigwedge B)\bigwedge C$
  associative laws\\
 3)$~~A\bigvee A = A, ~~A\bigwedge A = A$ idempotent laws\\
 4)$~~A = A\bigvee (A\bigwedge B), ~~A = A\bigwedge (A\bigvee B)$
 absorption laws\\
  By definition, a closed-set system is \emph{algebraic} if $\C$ is closed under unions
  of upward directed subsets, i.e., for every $S \subseteq \C$, $ ~\bigvee S \in \C $.
  Equivalently, the closure operator $\J$ of a closure-set system $<\Upsilon, \C >$ is algebraic
   if it satisfy the following "finitary" property: for any upward directed subset $X \subseteq \Upsilon$\\
 $~~~\J(X) = \bigcup\{ \J(X') ~|~~X'\subseteq_\omega X \}~$ \\
 where $ X'\subseteq_\omega X $ means that $X'$ is a \emph{finite }subset of $X$.\\
A lattice is \emph{algebraic} if it is \emph{complete} and
\emph{compactly generated}: a lattice $<\C, \subseteq >$ is
compactly generated if every element of $\C$ is a sup of compact
elements less then or equal to it, i.e., for every $A \in \C$, $ A =
sup\{ B\in Comp\C ~| ~B \subseteq A \}$  (an \emph{element is
compact} $B\in Comp\C$ if, for every $X \subseteq \C$ such that
$supX$ exists, $B\subseteq supX$  implies there exists a
$X'\subseteq_\omega X $ such that $ B\subseteq supX'$). Set of
compact elements in an algebraic lattice is the set of all closed
elements obtained from finite subsets.
\\ We define the \emph{finite objects} in $DB$ the databases
with a finite number of n-ary ($n$ is a finite number $ n \in
\omega$, the nullary relation is $\bot$ and is an element of each
object in $DB$ category) relations (elements); the \emph{extension}
of relations does not necessarily be finite - in such a case for a
finite object $A$ in $DB$, the object $TA$ is composed by infinite
number of relations, that is $TA$ is an infinite object.
 \\We will demonstrate that this
 database lattice is an \emph{algebraic lattice}.
 \begin{propo} \label{prop-closure} Let $\C = Ob_{DB_{sk}}$ denotes the set of all closed objects of DB category.
 The following properties for a database closure are valid:
\begin{itemize}
  \item  A closed-set system $<\Upsilon, \C >$ consists of the
  "total" closed object (top database instance) $\Upsilon \in \C$, which is  a merging (or, up to isomorphism, a union) of
  all objects in DB, and the set $\C \subseteq \P(\Upsilon)$, such
  that $\C$ is closed under intersections of arbitrary subsets.
  That is, for any  $K \subseteq \C$, $\bigcap K \in \C$.
  \item The closure operator $T$ is algebraic.
  \item $<\C, \subseteq >$ is an algebraic lattice with meet $\otimes$ and join $\oplus$ operators. The compact
  elements of $<\C, \subseteq >$ are closed objects of DB
  category $T(A)$ generated by  \emph{finite} objects $A\subseteq_\omega \Upsilon$.
\end{itemize}
 \end{propo}
\textbf{Proof:} It is easy to verify that $\C$ is closed under
intersection, it is  a poset of closed objects of  $DB$ category
(i.e., a set of objects of the equivalent skeletal category
$DB_{sk}$), which is a subset of the total object $\Upsilon$, with
set inclusion as  a partial ordering.\\
 The closure operator is $~T:Ob_{DB}\longrightarrow Ob_{DB}$.
 We have that each
 object $ A \in Ob_{DB}$ is a subset of $\Upsilon$, and
 vice versa, each subset of $\Upsilon$ is a database instance,
 thus an object in $DB$ category. From Universal algebra theory it holds that
 each closure operator, and its equivalent closure-set system $<\Upsilon, \C>$,  generates a complete lattice $<\C, \subseteq>$,
 such that for any subset $K\subseteq \C$ of closed objects $K
 = \{TA_i~|~i \in I, A_i~$closed set of$~DB\}$ we have that:\\
 Greatest lower bound $\bigwedge K = \bigwedge_{i\in I} TA_i =
   \bigcap _{i\in I}TA_i = \bigcap K = \otimes_{i\in I}TA_i$, that is, meet lattice operator $\bigwedge$ corresponds to
  the matching operation $\otimes$.\\
 Least upper bound $\bigvee K = \bigvee_{i\in I} TA_i   = T(\bigcup_{i\in
 I} A_i) =\oplus_{i\in I}TA_i$,that is, join lattice operator $\bigvee$ corresponds to
  the merging operation $\oplus$, \\ so that for $K =\C$ we obtain $\bigvee \C =  T(\bigcup_{A_i\in Ob_{DB}}A_i)
 =  \Upsilon$.\\
 Let us prove that $T$ is algebraic:
 let $\Upsilon_\Sigma = U(\Upsilon)$, where $~U:DB_{sk} \longrightarrow \K~$
 is the unique universal functor described previously in Section \ref{section:universal}, be a $\Sigma_R$-algebra
 generated by $\Upsilon$, and $A \subseteq \Upsilon$ a database
 instance (each object $A$ in $DB$ satisfy $A \subseteq \Upsilon$).
 The $A$ is subuniverse of $\Upsilon_\Sigma$ if for all
 $\Sigma_R$-algebra operators $\sigma \in \Sigma_R$, $ar(\sigma)
  = n$, and $a_1,...,a_n \in A$, $\sigma (a_1,...,a_n ) \in A$, i.e., $A$ is closed under $\sigma$ for each $\sigma \in
 \Sigma_R$. Thus, each subuniverse $A$ of $\Upsilon_\Sigma$ is a
 closed object in $DB$. The set of all subuniverses of
 $\Upsilon_\Sigma$ (i.e., the set of all closed objects of $DB$)
 is denoted by $Sub(\Upsilon_\Sigma)$.\\
 $\Upsilon_\Sigma$   defines, for every $A \subseteq
 \Upsilon$ the subuniverse \emph{generated by} $A$,\\ $Sg(A) = \bigcap\{ B | A \subseteq B~$and$~ B\in
 Sub(\Upsilon_\Sigma)\}$,\\ where $Sg:\P(\Upsilon)\longrightarrow
 Sub(\Upsilon_\Sigma)$ is an algebraic operator (Theorem
  of Universal
 algebra). Let us verify that $T \equiv Sg$:\\
 In fact, for any $A\in \P(\Upsilon)$, $A\in Ob_{DB}$, we obtain
 $Sg(A) = \bigcap\{ B ~|~ A \subseteq B~$ and$ ~ B\in
 Sub(\Upsilon_\Sigma)\} =  \bigcap\{ B ~| ~A \subseteq B~$ and $~
 B~$ is closed object in$ ~DB\} = \bigwedge\{ B ~| ~A \subseteq B~$and$~
 B~$ is closed object in$ ~DB\} = TA$ because $TA$ is the least
 closed object $B =TA$ in $DB$ such that $A\subseteq B$ (from the
 property of the closure operator $T$). Thus, $T$ is an algebraic
 closure operator and, consequently, the lattice $<\C, \subseteq>$ and the closed-set system $<\Upsilon, \C>$ are algebraic.
\\$\square$\\
 Now we can extend the lattice $<\C, \subseteq >$ of only closed
 objects of $DB$ into a lattice of all objects of $DB$ category:
\begin{propo}\label{prop-lattice} The set $Ob_{DB}$ of all database instances
(objects)of $DB$, together with merging and matching tensor products
$\oplus$ and $\otimes$ (read "join"  and "meet" respectively) is a
lattice.
\end{propo}
\textbf{Proof:} We have to prove that:\\
1)$~~A \oplus B \simeq B \oplus A, ~~A \otimes B \simeq B \otimes
A$,   commutative laws\\
 2)$~~A \oplus (B \oplus C) \simeq (A \oplus B) \oplus  A, ~~A \otimes(B\otimes C) \simeq (A\otimes B)\otimes
 C$,
  associative laws\\
 3)$~~A \oplus A \simeq A, ~~A \otimes A \simeq A$, idempotent laws\\
 4)$~~A \simeq A \oplus (A\otimes B), ~~A \simeq A\otimes (A \oplus
 B)$,
 absorption laws.\\
The commutative, associative and idempotent laws holds directly from
functorial definition of  $\oplus$ and $\otimes$. Let us prove (4):
We have that $ A \oplus (A\otimes B) = T(A \bigcup (TA \bigcap TA))
\subseteq T(A \bigcup TA) = TTA = TA$, thus we obtain that $ A
\oplus (A\otimes B) = TA \simeq A$.\\
 Analogously, $A\otimes (A \oplus B) = TA
\bigcap T(A \bigcup B) = TA \simeq A$.
\\$\square$\\
Let us denote by $T_I:DB_I \longrightarrow DB_I$ the restriction of
closure endofunctor $T:DB\longrightarrow DB$. We have seen in
Proposition \ref{prop:POCat} that $DB_I$ is a PO category where each
arrow $f:A\hookrightarrow B$ is a monomorphism, i.e., $A\preceq B$.
Thus, we obtained a partial order $<Ob_{DB}, \preceq>$. Let us show
that it is a lattice ordered set; i.e., that every pair of objects
$A,B \in Ob_{DB}$ has a least-upper-bound (sup) and the
greatest-lower-bound (inf).
\begin{propo}\label{prop:dblattice} Poset $<Ob_{DB}, \preceq >$ is a lattice
ordered set where $A\preceq B$ if $A \simeq A\otimes B$ (or
equivalently $B \simeq A\oplus B$), so that for all $A,B \in
Ob_{DB}$, $~inf(A, B)= A\otimes B$ and $sup(A, B) = A\oplus B$. It
is a complete lattice.
\end{propo}
\textbf{Proof:} In fact, if $A \simeq A\otimes B$ then $TA\bigcap TB
\simeq A \simeq TA$, thus $TA\bigcap TB = TA$ and $TA\subseteq TB$,
i.e., $A\preceq B$. Or, equivalently, if $B \simeq A\oplus B$ then
$TB = T(A \bigcup B) \supseteq TA$, thus $A\preceq B$. We have also
that $Ob_{DB}\supseteq \C$, where $<\C, \subseteq>$ is algebraic
lattice of closed objects. Thus, for any subset $K\subseteq Ob_{DB}$
we have that $inf(K) = \bigcap_{A_i\in K}TA_i~~\in~\C$, thus, from
$\C\subseteq Ob_{DB}$ we obtain that $inf(K) \in Ob_{DB}$, i.e., the
lattice $<Ob_{DB}, \preceq >$ for every subset $K$ has a
least-upper-bound and, consequently, it is a complete lattice.
\\$\square$
\begin{coro} \label{coro:algebr_lattice} PO subcategory $DB_I \subseteq DB$, $DB_I = <Ob_{DB}, \preceq >$ is an algebraic
lattice isomorphic to the lattice $<\C , \subseteq>$.
\end{coro}
\textbf{Proof:} If we define an equivalence classes for $<Ob_{DB},
\preceq>$ w.r.t. the equivalence relation "$\simeq$", such that $[A]
= \{ B~|~B\in Ob_{DB} ~$and$~B \simeq A  \}$, so that $<\C,
\subseteq>$ is its quotient lattice (we consider latices as
algebras) such that elements of this quotient lattice (algebra) are
 closed objects $[A] = TA$ only. The function $\alpha:<Ob_{DB},
\preceq> \longrightarrow \\ <\C, \subseteq >$, such that for any
$A\in Ob_{DB}$, $\alpha(A) = \alpha(TA)= TA$, i.e., $\alpha \equiv
T$, is an order-preserving bijection ($A$ and $TA$ are
indistinguishable elements in the lattice $<Ob_{DB}, \preceq >$,
thus $|<Ob_{DB}, \preceq >| = |<\C, \subseteq >|$), while the
function $\alpha^{-1}:<\C, \subseteq > \longrightarrow <Ob_{DB},
\preceq >$ is an order-preserving identity function. Thus, $\alpha $
is an isomorphism of lattices, and, consequently, also $<Ob_{DB},
\preceq
>$ is algebraic.
\\$\square$\\
Database lattice $DB_I = <Ob_{DB}, \preceq >$  is \emph{bounded}: it
 has the largest element $\Upsilon$ (element that is upper bound of
 every element of the lattice), and also the smallest element
 $\perp^0$. The algebraic property is very useful in order to
 demonstrate the properties of $DB$ category: in order to prove theorems
 in general we need to be able to extend inductive process of
 proof beyond $\omega$ steps to the transfinite. \emph{Zorn's lemma}
 (equivalent to the Axiom of Choice of set theory) allows us to do
 this. The database lattice $<Ob_{DB}, \preceq >$ is a (nonempty) poset with the property that every chain
  $K \subseteq Ob_{DB}$ (i.e., linearly ordered
 subset) has an upper bound $\bigvee K = \bigcup K$ (because
 this  poset is \emph{algebraic}) in $Ob_{DB}$. Then we can apply the Zorn's lemma
 which asserts that $<Ob_{DB}, \preceq >$ has a maximal element.\\
 \textbf{Remark:} From the fact that $DB_I = <Ob_{DB}, \preceq >$  is an
algebraic lattice we obtain that for the total object $\Upsilon$ the
following is valid: $~~~\Upsilon = T\Upsilon = \bigcup \{ TA~ |~A
\subseteq_\omega \Upsilon \} = \bigvee \{TA ~|~A ~$ is finite such
that $A\preceq \Upsilon \}$, it is the union of all closed objects
generated by only \emph{finite} objects of $DB$, i.e., the union of
all \emph{compact}  elements of $<\C, \subseteq
>$. Similarly, in an algebraic lattice every element is generated as
lub of the set of compact elements which are less that this element,
that is we have that $A \simeq \bigvee \{TB ~|~B ~$ is finite such
that $B\preceq A \}$. Each closed
object $TA$ which is not compact object is obtained from an infinite object (database) $A$.\\
 Let $\omega$ be the category of natural numbers with arrows
 $\leq:j\longrightarrow k$ which correspond to the total order
 relation $j\leq k$, i.e., $\omega = \{ 0 \rightarrow 1\rightarrow 2\rightarrow
 ....\}$.  An endofunctor $H:C \longrightarrow D$ is \emph{$\omega-cocontinuous$} if preserves the
 colimits of functors $J:\omega \longrightarrow C$, that is when $H
 ColimJ \simeq ColimHJ$ (the categories $C$ and $D$ are thus supposed
 to have these colimits). Notice that a functor $J:\omega \longrightarrow
 C$ is a diagram in $C$ of the form $\{ C_0 \rightarrow C_1\rightarrow C_2\rightarrow
 ....\}$. For $~\omega-cocontinuous$ endofunctors the construction
 of the \emph{initial algebra} is inductive ~\cite{SmPl82} .
 \begin{propo} \label{prop:cocontinuity} For each object $A$ in the category $DB$ the"merging with $A$"  endofunctor $\sum_A = A\oplus \_:DB\longrightarrow DB$ is
 $\omega-cocontinuous$.
 \end{propo}
 \textbf{Proof:} Let us consider any chain in $DB$ (all arrows are
 monomorphisms, i.e., "$\preceq$" in a correspondent chain of the $<Ob_{DB}, \preceq >$ algebraic
 lattice), is a following diagram $\D$\\
 $  \perp^0 ~\preceq_0 ~(\sum_A \perp^0) ~\preceq_1~ (\sum^2_A \perp^0)~ \preceq_2~~~
 ... ~~~~~~\sum^{\omega}_A  $,\\
  where $\perp^0$ is the initial object in $DB$, with unique monic arrow $\perp = \preceq_0:\perp^0 \hookrightarrow (\sum_A
  \perp^0)$ with $\widetilde{\perp} = \perp^0$,
  and consecutive arrows $\preceq_n = \sum^n_A \perp: (\sum^n_A \perp^0) \hookrightarrow
  (\sum^{n+1}_A \perp^0)$ with  $\widetilde{\sum^n_A \perp} = TA$, for all $n \geq 1$,
 as representation of a
 functor $J:\omega \longrightarrow DB$. The endofunctor $\sum_A$
 preserves colimits because it is monotone and $\sum^{\omega}_A = TA$ is its
 fixed point, i.e., $\sum^{\omega}_A = TA  = T(A \bigcup TA) = T(A \bigcup \sum^{\omega}_A) = \sum_A(\sum^{\omega}_A)$.
 Thus, the colimit   $ColimJ = \sum^{\omega}_A$  of
 the base diagram $\D$ given by the functor $J:\omega \longrightarrow DB$, is equal to $ColimJ = (A\oplus \_)^{\omega}\perp^0 = TA$.
 Thus $\sum_AColimJ = T(A \bigcup ColimJ) = T(A\bigcup TA) = T(TA) = TA = Colim\sum_AJ$ (where $Colim\sum_AJ$ is a colimit of the diagram
 $\sum_A \D$).\\
  The $\omega-cocompleteness$ amounts to chain-completeness,
 i.e., to the existence of least upper bound of $\omega-chains$.
 Thus $\sum_A$ is $\omega-cocontinuous$ endofunctor: a monotone
 function which preserves lubs of $\omega-chains$.
 \\$\square$\\
 In what follows we will pass from lattice based
 concepts, as lubs of directed subsets, compact subsets, and algebraic
 lattices, to categorially generalized  concepts as directed
 colimits, finitely presentable (fp) objects, and locally finitely
 presentable (lfp)m categories respectively:\\
 A \emph{directed colimit} in $DB$ is
 a colimit of the functor $F:<J,\preceq> \longrightarrow DB$, where $<J,\preceq>$ is
 a directed partially ordered set, such that for any two objects $j, k \in J$ there is an object $l \in J$
 such that $j\preceq l, k\preceq l$, considered as a category. For example, when $J =
 Ob_{DB}$ we obtain the algebraic (complete and compact) lattice
 which is an directed PO-set, such that for any two objects $A,B \in
 J$ there is an object $C \in J$ with $A\preceq C$ and $B\preceq
 C$ (when $C = sup(A,B)\in J$).\\
 An object $A$ is said to be \emph{finitely presentable} (fp), or finitary, if
 the functor $DB(A,\_):DB \rightarrow Set$ preserves directed colimits (or, equivalently, if it preserves filtered
 colimits).
  We write $DB_{fp}$ for the full subcategory of $DB$ on the finitely
 presentable objects: it is essentially small. Intuitively, fp objects are "finite objects",
  and a category is lfp if it can be generated from its  finite objects:
  a \emph{strong generator} $M$ of a category  is its small full
  subcategory such that $f:A\longrightarrow B$ is an isomorphism
  \emph{iff} for all objects $C$ of this subcategory, given a
  hom-functor $M(C,\_):M\longrightarrow Set$, the following
  isomorphism of hom-setts $M(C,f):M(C,A) \longrightarrow M(C,B)$ in $Set$ is
  valid.\\
  From Th.1.11 ~\cite{AdRo94} a category is \emph{locally finitely
  presentable} (lfp)  iff it is cocomplete and has a strong generator.
\begin{coro} \label{coro:lfp} $DB$ and $DB_{sk}$ are concrete, locally small,
and locally finitely presentable categories (lfp).
\end{coro}
\textbf{Proof:} Given any two objects $A,B$ in $DB$, the hom-set
$DB(A,B)$ of all arrows  $f:A\longrightarrow B$ corresponds to the
directed subset
 $K = \{ \widetilde{f} ~|~\perp^0 \subseteq \widetilde{f}\subseteq A\otimes
 B\} \subseteq \C$, which is bounded algebraic (complete and compact) sublatice of $\C$.
  Thus, the set of all arrows $f:\Upsilon\longrightarrow
 \Upsilon$ corresponds to the directed set $K = \{ \widetilde{f} ~|~\perp^0 \subseteq
 \widetilde{f}\subseteq \Upsilon\otimes \Upsilon = \Upsilon \}$,
 which is equal to the lattice $<\C, \subseteq>$. Thus, $DB$ is \emph{locally small}
 (has small hom-sets), and, by $DB \supseteq DB_{sk}$, also $DB_{sk}$ is locally
 small.\\
 Let us show that the full subcategory $DB_{fin}$, composed by closed objects obtained from  \emph{finite} database objects,
 is a strong generator of $DB$: in
  fact, if $TA,TB \in DB$ and $A\simeq B$ are two \emph{finite} databases (so that $TA = TB$) then for all $C \in DB_{fin}$, $|DB(C,TA)|$ is a
  rank of the complete sublattice $<Ob_{DB}, \preceq>$ bounded
  by $\perp^0 \preceq D \preceq C\otimes TA$, while $|DB(C,TB)|$ is a
  rank of the complete sublattice $<Ob_{DB}, \preceq>$ bounded
  by $\perp^0 \preceq D_1 \preceq C\otimes TB$. From $A\simeq B$
  we deduce $C\otimes TA = TC \bigcap TA = TC\bigcap TB = C\otimes
  TB$, thus $|DB(C,TA)| = |DB(C,TB)|$, i.e, there is a bijection $\upsilon:|DB(C,TA)| \simeq
  |DB(C,TB)|$ which is an isomorphism in $Set$. Thus,
  $DB$, which is cocomplete and has this strong generator $DB_{fin}$, is a lfp.
\\$\square$\\
 We define a representable functor $DB(A,\_):DB \longrightarrow
 Set$, such that $DB(A,B)$ is the set of functions $\{ F_{DB}(f)~|~$for each$~f:A\longrightarrow B ~$ in$ ~DB \}$, and for
 any arrow $g:B\longrightarrow C$, $DB(A, g)$ is the function
 such that for any function $f \in DB(A,B)$ we obtain the
 function $h = DB(A,g) \triangleq F_{DB}(g) \circ f \in
 DB(A,C)$.\\
 We say that a functor $H:DB\longrightarrow Set$ preserves
 colimits if the image $H\nu : HF \longrightarrow H ColimF$ for
 the colimit $(\nu, ColimF)$ of a functor $F \in DB^J$  is a
 colimiting cone (or cocone) for $HF$ (in this case we are interested for $H = DB(\Upsilon,
 \_)$).\\
 Let us show, for example, that the object $\Upsilon$ is a finitely presentable
 (fp) (it was demonstrated previously by remark that $~~~\Upsilon = T\Upsilon = \bigcup \{ TA~ |~A
\subseteq_\omega \Upsilon \} = \bigvee \{TA ~|~A ~$ is finite such
that $A\preceq \Upsilon \}$), i.e., the fact that its hom-functor
$DB(\Upsilon, \_):DB \longrightarrow
 Set$ preserves directed colimits:
\begin{propo} \label{prop:fp} Total object (matching monoidal unit) $\Upsilon$ is
a finitely presentable (fp).
\end{propo}
 \textbf{Proof:} Let us have a $ColimF$ in $DB$
 (a colimit of the functor $F \in DB^J$, where $F$ can be seen as a
 base diagram for this colimit, composed by a finite number of
 objects $B_1,....,B_n$ with PO-arrows "$\preceq $" between them),
 such that arrows $h_i:B_i \hookrightarrow ColimF$ are components
 of the cone $(\nu, ColimF)$ where $\nu:F \longrightarrow \triangle
 ColimF$ is a natural transformation and $\triangle $ is a
 diagonal constant functor. \\
 Let us show that for any other cocone $E$ in $Set$, for the same cocone-base
 $HF$ (where $H = DB(\Upsilon,\_):DB \rightarrow Set$) there is an unique arrow (function) from $DB(\Upsilon
 ,ColimF)$ to the set $E$ (vertex of a cocone $E$). We can see that, for a set of all objects in the diagram (functor) $F$,
  $S = \{~B_i~|~B_i \in F \} \subseteq Ob_{DB}$,
  holds that $ColimF = sup(S) = sup \{~B_i~|~B_i \in F \} =  \sum_{B_i \in F} B_i$.  Each hom-set $DB(\Upsilon,
 B_i)$ in $Set$ is isomorphic to the complete sublattice
  of the algebraic lattice $<\C,\subseteq >$,  $<\{\widetilde{f}|\widetilde{f} \subseteq TB_i\}, \subseteq >$
   (because each arrow $f:\Upsilon \longrightarrow B_i$
corresponds to the  closed object $\widetilde{f} \subseteq \Upsilon
\otimes B_i = TB_i$). On the other hand
   $HColimF = DB(\Upsilon , ColimF)$ is isomorphic to the
 complete sublattice $<S, \subseteq>$, where $S =\{\widetilde{f}|\widetilde{f} \subseteq \sum_{B_i \in F} TB_i\}$.
   Thus, all arrows of  the cocone $H\nu$, $DB(\Upsilon, h_i): DB(\Upsilon, B_i) \hookrightarrow DB(\Upsilon, ColimF)$ are
 inclusion functions $<\{\widetilde{f}|\widetilde{f} \subseteq TB_i\}, \subseteq> ~ \subseteq ~ <S, \subseteq>$ (also each arrow in
 the base diagram $HF$ in $Set$ are inclusion functions $<\{\widetilde{f}|\widetilde{f} \subseteq TB_j\}, \subseteq>~ \subseteq
  ~< \{\widetilde{f}|\widetilde{f} \subseteq TB_k\}, \subseteq>$).\\
   All arrows of the cocone $E$, $~k_i:< \{\widetilde{f}|\widetilde{f} \subseteq TB_i\}, \subseteq>\longrightarrow E$,
    must be equal functions (only with different domains) in order
 to preserve the commutativity of this colimiting cocone $E$: thus the function $k:<S, \subseteq> \longrightarrow
 E$ is an unique function such that, for any $v \in <S, \subseteq>$, $k(v)
 = k_i(v)$ for some $k_i:<\{\widetilde{f}|\widetilde{f} \subseteq TB_i\}, \subseteq> \longrightarrow E$
 and $ v\in <\{\widetilde{f}|\widetilde{f} \subseteq TB_i\}, \subseteq>$.  From $HColim = DB(\Upsilon, ColimF) \simeq
  S$ we conclude that there is an unique arrow in $Set$ from
  $HColimF$ to $E$. Thus, $HColim$ is a colimit in $Set$, i.e., $H = DB(\Upsilon,
  \_)$ preserves directed colimits and, consequently, $\Upsilon$
  is a finitely presentable.
 \\$\square$\\
 \textbf{Remark:} We emphasize the fact that $\Upsilon$ is fp object for a
 more general considerations of the theory of enriched categories, which will be elaborated in Section
 \label{section:enricmnet}, as demonstration that the monad based on
 the power-view endofunctor $T:DB\rightarrow DB$ is an enriched
 monad. The Kelly-Power theory applies in the case of a symmetric
 monoidal closed category, which is lfp and closed category (which
 is equivalent to demanding that the underlying ordinary category is
 lfp, and that the monoidal structure on this ordinary category
 restricts to one on its fp objects. For details see \cite{Kell82,KePo93}, but
 note in particular that the unit $\Upsilon$ must be finitely
 presentable.\\
 A locally finitely presentable category ~\cite{GaUl71} is the category of models
 for an essentially \emph{algebraic} theory, which allows
 operations whose domain is an equationally defined subset of some
 product of the previously defined domains (the canonical example
 is a composition in a category, which is defined only on
 composable, not arbitrary pairs of arrows). In fact, we deduce
 from the algebraic (complete and compact) lattice $<Ob_{DB}, \preceq >$  that for any object $A$ holds that $A \simeq TA = \oplus \{
 TB~|~B\subseteq_\omega A \} = \oplus S$ (remember that $\oplus$ is a generalization in $DB$ of the union operation $\bigcup$ for sets and $X \oplus Y = TX \oplus TY$),
 where the set $S = \{B~|~B\subseteq_\omega A  \}$ is upward directed, i.e., for any two finite $B_1, B_2\subseteq_\omega A$ there is $C = B_1 \bigcup B_2\in S$ such that
 $B_1 \preceq C, B_2 \preceq C$, with $TC = B_1 \oplus B_2$), i.e., any object in $DB$ is
 generated from finite objects and this generated object is just a
 directed colimit of these fp $~$ objects.\\
 An important
 consequence of this freedom is that we can express \emph{conditional}
 equations in the logic for databases.\\
 Other important result from the fact that $DB$ is a complete and cocomplete lfp category
 that it can be used as the category of models for essentially
 algebraic theory \cite{Lawve63,Powe00} as is a relational database theory. Thus it is a
 category of models for a finite limit sketch, where sketches are
 called graph-based logic \cite{KiPT99,BaWe85}, and it is well known that a
 relational database \emph{scheme} can readily be viewed, with some
 inessential abstraction involved, as a sketch. By Liar's theorem, a
 category $DB$ is accessible \cite{Liar96,Mapa89}, because it is
 sketchable.\\
 \textbf{Remark:} differently from standard application of sketches used to define a theory of a
 \emph{single} database scheme, so that objects of this graph-based logic
 theory are single relations of such a database and arrows between
 them are used to define the common database functional
 dependencies, inclusion dependencies and other database
 constraints, in the case of inter-database mappings we need to use
 the whole databases as objects in this lfp $DB$ category: the price
 for this higher level of abstraction  is that arrows in $DB$ are
 much more complex than i standard setting and that generally are
 not functions.
 \section{Enrichment \label{section:enricmnet}}
 It is not misleading, at least initially, to think of an enriched
 category as being a category in which the hom-sets carry some
 extra structure (partial order $\preceq$ of algebraic sublattice $<Ob_{DB}, \preceq>$ in our case) and in which that structure is preserved by
 composition. The notion of enriched category ~\cite{Kell82} is
 more general and allows for the hom-objects ("hom-sets") of the
 enriched category to be objects of some monoidal category,
 traditional called $V$.\\
 Let us now prove that $DB$ is a \emph{monoidal closed}
 category: for any two objects $B$ and $C$  the set of all arrows $\{f_1,f_2,...\}:B\rightarrow  C$, from $B$ into $C$, can be represented by an  unique arrow
 $(\bigcup_{f_i \in DB(B,C)} f_i):B \rightarrow C$, so that the object $C^B$ is equal to the information flux of this arrow $\widetilde{\bigcup_{f_i \in DB(B,C)} f_i}$
 . Thus, we define the
 hom-object $C^B \triangleq \oplus_{f \in
 DB(B,C)} \widetilde{f}~~$ (\emph{merging} of all closed objects obtained from a hom-set of
 arrows from $B$ to $C$), i.e., merging of compact elements $A\preceq B\otimes C$ (where $B\otimes C$ is the "distance" between $B$ and $C$,
 following Lawvere's idea, as follows from the definition of metric space for $DB$ category in a Section \ref{Sect:metric})
  which "internalize" the hom-sets.\\
  Thus we obtain that $C^B = T(\bigcup \{ \widetilde{f}~|~f \in
 DB(B,C)\}) = T(\bigcup \{ \widetilde{f}~|~\widetilde{f} \subseteq
 B\otimes C)\}) = T(\bigcup \{ \widetilde{f}~|~\widetilde{f} \preceq
 B\otimes C)\}) = T(B\otimes C) = B\otimes C$.\\
 Generally a monoid $M$ acting on set $Ob_{DB}$ may be seen as
 general metric space where for any $B \in Ob_{DB}$ the distance $C^B
 $ is a set of $v\in S$ (views on our case) whose action send $B$
 to $C$ (gives a possibility to pass from the "state" $B$ to
 "state $C$ of the database "system" of objects in $DB$).\\
 A monoidal category is closed if the functor $\_ \otimes B :DB \longrightarrow DB$
 has a right adjoint $(\_~)^B:DB \longrightarrow DB$ for every object
 $B$ , $< (\_~)^B ,~ \_ \otimes B ,~ \eta_\otimes , \varepsilon_\otimes >:DB \longrightarrow DB$ , with the counit $\varepsilon_C:C^B \otimes B\longrightarrow
 C$ called the evaluation at $C$ (denoted by
 $eval_{B,C}$).
 \begin{propo} \label{prop:moncat1} Strictly symmetric idempotent monoidal
 category $(DB,\otimes ,\Upsilon )$ is a \textsl{monoidal bi-closed}:
 for every object $B$ , there exists an isomorphism $\Lambda:
 DB(A\otimes B ,C)~\simeq~DB(A , C^B)~~$ such that for any $f\in
 DB(A\otimes B ,C)~$ , $\Lambda(f) \approx f~$ , the hom-object
 $C^B$ together with a monomorphism $eval_{B,C}:C^B\otimes B\hookrightarrow
C$ the following "exponent" diagram
\begin{diagram}
   C^B\otimes B  & \rInto^{eval_{B,C}}   & C \\
 \uTo^{\Lambda(f)} \uTo_{id_B}& \ruTo_{f} &\\
 A\otimes B  &      &     \\
 \end{diagram}
commutes, with
$~~~~~~f = eval_{B,C}\circ ( \Lambda(f) \otimes id_B)~$.
\end{propo}
\textbf{Proof:} From a definition of hom-object we have $C^B
\triangleq T(\bigcup \{ \widetilde{g}~|~g \in
 DB(B,C) \}) = T(\bigcup \{ \widetilde{g}~|~ \perp^0 \subseteq \widetilde{g} \subseteq B\otimes C \}) = B\otimes
 C$. Thus, from $ \widetilde{f} \subseteq A\otimes B \otimes C = TA \bigcap TB
 \bigcap TC$, and the fact that for a monomorphism  $\widetilde{eval_{B,C}} = T(C^B\otimes B) = T(B \otimes C) = B \otimes C$,  we obtain for the commutativity of
 this exponential diagram that,
  $ \widetilde{f} = \widetilde{eval_{B,C}\circ ( \Lambda(f) \otimes id_B)}\\ =
 \widetilde{eval_{B,C}} \bigcap \widetilde{( \Lambda(f) \otimes
id_B)} = \widetilde{eval_{B,C}} \bigcap (\widetilde{ \Lambda(f)}
\otimes \widetilde{id_B} ) = B\otimes C \bigcap \widetilde{
\Lambda(f)}\bigcap TB = TB \bigcap TC \bigcap \widetilde{
\Lambda(f)} = \widetilde{ \Lambda(f)}$, from the fact that
$\widetilde{ \Lambda(f)} \subseteq C^B = TB \bigcap TC$.
\\Thus, $~~f =
eval_{B,C}\circ ( \Lambda(f) \times id_B)~$ iff $\Lambda(f) \approx f~$.\\
$\Lambda~$ is a bijection, because $DB(A\otimes B , C) = \{ g~|~
\perp^0 \subseteq \widetilde{g} \subseteq A\otimes B\otimes C \}
\cong \{ \widetilde{g}~|~ \widetilde{g}\in K \}$, where $K$ is a
bounded algebraic sublattice (of closed objects) of the lattice
$(\C, \subseteq )$ and $\cong$ denotes a bijection, i.e., $K = \{
a~|~ a \in \C ~and~a\subseteq TA\bigcap TB\bigcap TC \}$. Also $DB(A
, C^B ) = DB( A , B\otimes C) \cong K$ , thus $|DB(A\otimes B, C )|=
|DB(A, C^B )| = |K|$, Thus,  $\Lambda~$ is a bijection, such that
for any $f \in DB(A\otimes B, C )$ , $\widetilde{\Lambda (f) } =
\widetilde{f} \in K~$, i.e., $\Lambda (f) \approx f$.\\
Consequently, $DB$  is closed and symmetric, that is, biclosed
category.
\\$\square$\\
 \textbf{Remark:} from duality we have that, for any two objects $A$ and $B$
 that $|DB(A, B )|= |DB(B, A )|~$ , i.e., $A^B = B^A \equiv A\otimes
 B$. That is, the \emph{cotensor} (hom object) of any two objects $A^B$ which is a particular limit in $DB$
  is equal to the correspondent colimit of these two object, that is \emph{tensor product}  $A\otimes
 B$: this fact is based on the duality property of $DB$ category. \\
 We have seen that all objects in $DB$ are finitely representable.
 Let us denote by $V = DB(\Upsilon, \_):DB \longrightarrow Set$ the
 representable functor $DB(\Upsilon, \_)$. By putting $A = \Upsilon $
 in $\Lambda$, and by using the isomorphism $\beta:\Upsilon\otimes B
 \simeq B$, we get a natural isomorphism $DB(B,C) \simeq V(C^B) =
 DB(\Upsilon, C^B)$. Than $C^B$ is  exhibited as a lifting
 through $V$ of the hom-set $DB(B,C)$, i.e., hom-object $C^B$ is a
 set of all views which gives a possibility to pass from a
 "state" $B$ to a "state" $C$. It is called the \emph{internal hom} of $B$  and $C$.\\
 By putting $B = \Upsilon $ in $\Lambda$ and by using the isomorphism
 $\gamma:A \otimes \Upsilon \simeq A$ we deduce a natural
 isomorphism $i:C \simeq C^\Upsilon $ (it is obvious by
 $C^\Upsilon = C \otimes \Upsilon \simeq C)$.\\
 The fact that a monoidal structure is closed means that we have an internal Hom functor, $(\_~)^{(\_~)}:DB^{op}\times DB\rightarrow
 DB$, which 'internalizes' the external Hom functor, $Hom:DB^{op}\times
 DB\rightarrow Set$, such that for any two objects $A,B$, the
 hom-object $B^A = (\_~)^{(\_~)}(A,B)$, represents the hom-set $Hom(A,B)$
 (the set of all morphisms from $A$ to $B$).\\
 We have that $(A\oplus B)\otimes C = (TA \bigcup TB)\bigcap TC = (TA
 \bigcap TC) \bigcup (TB \bigcap TC) = (A \otimes C) \bigcup (B\otimes C)
  \simeq (A \otimes C) \oplus (B\otimes C)$ and $C\otimes (A\oplus B) = TC \bigcap (TA \bigcup TB) = (TC \bigcap TA)\bigcup (TC \bigcap TB)
  \simeq
 (C \otimes A) \oplus (C\otimes B)$, and $A\otimes \perp^0 \simeq \perp^0 \simeq
 \perp^0 \otimes A$ for the initial object $\perp^0$.\\
  Monoidal closed categories generalize Cartesian closed ones in
 that they also posses exponent objects $B^A$ which "internalize"
 the hom-sets. One may then ask if there is a way to
 "internally" describe the behavior of functors on morphisms . That is,
 given a monoidal closed category $C$ and a functor $F:C\longrightarrow
 C$ , consider, say, $f\in C(A,B)$ then $F(f)\in C(F(A), F(B))$ .
 Since hom-object $B^A$ and $F(B)^{F(A)}$ represent hom-sets $C(A,B)$ and
 $C(F(A) , F(B))$ in $C$ , one may study the conditions under
 which $F$ is "represented" by morphism in $C(B^A , F(B)^{F(A)})$ ,
 for each $A$ and $B$.
 \begin{propo} \label{prop:v-category} The endofunctor $T:DB\longrightarrow DB$ is \textsl{closed}.\\
 DB is a V-category \textsl{enriched over itself}, with the
 composition law monomorphism $m_{A,B,C}:C^B \otimes B^A \hookrightarrow
 C^A $ and identity element (epimorphism) $j_A:\Upsilon \twoheadrightarrow
 A^A$ which "picks up" the identity in $A^A$.\\
 The monad $(T, \eta, \mu )$ is an \textsl{enriched monad}, thus,
 $DB$ is an object of V-cat, and  endofunctor $T:DB\rightarrow DB$ is an
 arrow of V-cat.
 \end{propo}
 \textbf{Proof:} It is easy to verify that for each two objects (databases) $A$ and $B$ in
 $DB$ there exists $f_{AB}\in DB(B^A , (TB)^{TA})$, called "an \emph{action} of $T$ on $B^A$, such that for
 all $g\in DB(A,B)$ is valid\\
 $f_{AB}\circ \Lambda (g \circ \beta(A)) = \Lambda (T(g)\circ
 \beta(TA)): \Upsilon \longrightarrow (TB)^{TA}$ , where $\beta:\otimes \_ \longrightarrow I_{DB} $
 is a left identity natural transformation of a monoid $(DB, \otimes, \Upsilon, \alpha, \beta,
 \gamma)$, thus $ \beta (A) = id_A$, $\beta (TA) = id_{TA}$. In
 fact, we take $f_{AB} = id_{B^A}$, and we obtain,\\ $\widetilde{f_{AB}\circ \Lambda (g \circ
 \beta(A))}= \widetilde{id_{B^A}\circ \Lambda (g \circ id_A)} =
  \widetilde{\Lambda(g)} = \widetilde{g} = \widetilde{T(g)}=
  \widetilde{\Lambda(T(g))} = \widetilde{\Lambda(T(g)\circ
  id_{TA})}= \widetilde{\Lambda(T(g)\circ
  \beta (TA) )}$.\\ Consequently, $T$ is a closed endofunctor.\\
  The composition law $m_{A,B,C}$ may be equivalently represented by
  a natural transformation $m:(B\otimes \_ )\otimes (\_\otimes B)
  \longrightarrow \otimes $, and  an identity element $j_A$ by
  natural transformation $j:Y\longrightarrow \otimes\circ
  \bigtriangleup $, where $\bigtriangleup:DB\longrightarrow
  DB\times DB$ is a diagonal functor, while $Y:DB\longrightarrow
  DB$ is a constant endofunctor, $Y(A) \triangleq \Upsilon $ for
  any $A$ and $Y(f) \triangleq id_\Upsilon $ for any arrow $f$ in
  $DB$. It is easy to verify that two coherent diagrams
  (associativity and unit axioms) commute, thus $DB$ is enriched
  over itself V-category (as, for example, $Set$ category).\\
  $T$ is a V-functor: for each pair of objects $A, B$ there exists
  an identity map (see above) $f_{AB}:B^A \longrightarrow (TB)^{TA}$, subject to the compatibility with composition $m$ and with the
  identities expressed by the commutativity $f_{AB}\circ m_{A,B,C}
  = m_{TA,TB,TC}\circ (f_{AB}\otimes f_{AB})$ and $j_{TA} =
  f_{AB}\circ j_A$.  It is easy to verify that also natural
  transformations $\eta:I_{DB}\longrightarrow T$, $\mu:T^2\longrightarrow
  T$ satisfy the V-naturality condition (V-natural transformation
  $\eta$ and $\mu$ are an $Ob_{DB}$-indexed family of components
  $\delta_A:\Upsilon \twoheadrightarrow TA$ in $DB$ (for $\eta$, $\delta_A:\Upsilon \twoheadrightarrow
  (TA)^A$, $(TA)^A = TA$; while for $\mu$, $\delta_A:\Upsilon \twoheadrightarrow
  TA^{T^2A}$, $(TA)^{T^2A} = TA$). This map $f_{AB}$ is equal also
  for the endofunctor identity $I_{DB}$, and for the endofunctor
  $T^2$, because $B^A = B\otimes A = (TB)^{TA} = (T^2B)^{T^2A}$.
 \\$\square$\\
  In fact, each monoidal closed category is itself a V-category:
  hom-sets from $A$ to $B$ are defined as "internalized"
  hom-objects (cotensors) $B^A$. The composition is given by the image of the
  bijection $\Lambda:DB(D\otimes A, C)\simeq DB(D, C^A)$,
  where $D = C^B\otimes B^A$, of the arrow  $\varepsilon_B \circ
  (id_{C^B}\otimes \varepsilon_A )\circ \alpha_{C^B,B^A,A}$, i.e., $m_{A,B,C} = \Lambda(\varepsilon_B \circ
  (id_{C^B}\otimes \varepsilon_A )\circ \alpha_{C^B,B^A,A}) = \Lambda(eval_{B,C} \circ
  (id_{C^B}\otimes eval_{A,B})\circ \alpha_{C^B,B^A,A})$ (it is a monomorphism,
  in fact, $~\widetilde{m_{A,B,C}} =  \widetilde{\Lambda(eval_{B,C} \circ
  (id_{C^B}\otimes eval_{A,B})\circ \alpha_{C^B,B^A,A})} = TA\bigcap TB \bigcap TC =
  T(C^B \otimes B^A)$). The identities
  are given by the image of the isomorphism
  $\beta_A:\Upsilon\otimes A \longrightarrow A$, under the
  bijection $\Lambda:DB(\Upsilon\otimes A, A)\simeq DB(\Upsilon,
  A^A)$ , i.e., $j_A = \Lambda (\beta_A):\Upsilon
  \twoheadrightarrow A^A$ ($j_A$ is an epimorphism because, $\widetilde{j_A} = \widetilde{\Lambda (\beta_A)}
  = \widetilde{\beta_A} =TA = T(A^A)$).\\
  Moreover, for a V-category $DB$ holds the following isomorphism
  (which extends the tensor-cotensor isomorphism $\Lambda$ of
  exponential diagram in Proposition \ref{prop:moncat1})
  valid in all enriched Lawvere theories \cite{Powe00}, $DB(D\otimes A, C)\simeq DB(D,
  C^A) \simeq DB(A,DB(D,C))$.\\
  Finaly, from the fact that $DB$ is a lfp category enriched over
  the lfp symmetric monoidal closed category with a tensor product
  $\otimes$ (matching operator for databases), and the fact that $T$
  is a finitary enriched monad on $DB$, by Kelly-Power theorem we
  have that $DB$ admits a presentation by operations and equations,
  what was implicitelly assumed in the definition of this power-view
  operator in \cite{Majk04AOT,Majk08db}.
\section{Topological properties}
In this Section we will investigate some topological properties of
database category $DB$. That is we will consider its metric,
subobject classifier and topos properties. \\
We will show that $DB$
is a metric space, weak monoidal topos and some negative results as:
it is not well-pointed, has no power objects and pullbacks does not
preserve epics.
\subsection{Database metric space \label{Sect:metric}}
In a metric space $X$, we denote by $X(A,B)$ the non negative real
quantity of X-distance from the point $A$ to the point $B$. In a
database context, for any two given databases $A$ and $B$, their
matching is inverse proportional to their distance: The maximal
distance, $\infty$, between any two objects is equal to the minimal
possible matching, i.e., $\infty$ is represented by the closed
object $\bot^0$, while the minimal distance, $0$, we obtain for
their maximal matching, i.e., when these two objects are
isomorphic ($A \simeq B$).\\
Following this reasoning, we are able to define formally the concept
of the database distance:
\begin{definition}\label{def:distance} If $A$ and $B$ are any two objects in $DB$,
then their distance, denoted by $~d(A,B)$, is defined as follows:\\
\begin{displaymath}
   d(A,B) = \left\{ \begin{array}{ll}
   \Upsilon & \textrm{ ~, ~if ~$ A \simeq B$}\\
  A^B & \textrm{~,~othervise}
\end{array} \right.
\end{displaymath}
The (binary) partial \emph{distance relation} $\sqsubseteq$, on
\textsl{closed} database objects, is defined as inverse of the set
inclusion relation $\subseteq$.
\end{definition}
Notice that each \emph{distance} is a \emph{closed database object}
(such that $A = T(A)$): the minimal distance $\Upsilon$ (total
object), the maximal distance $\bot^0$ (zero object), and
hom-objects $B^A$ ($B^A = T(A) \bigcap T(B)$, intersection of two
closed objects is a closed object also).\\
Thus, a database metric space $DB_{met}$, where points are databases
and their distances are \emph{closed }databases, is a subcategory of
$DB$, composed by only epimorphic arrows: each epimorphism $f:A
\rightarrow B$ (i.e., $A \supseteq B$) in $DB_{sk}$, correspond to
the distance relation $A\sqsubseteq B$. Thus we can say that a
database metric space is embedded in $DB$ category, where distances
are closed databases and distance
relations are epimorphisms between closed databases.\\
Let us show  that this definition of distance for databases
satisfies
the general metric space properties.\\
A categorical version of metric space under the name enriched
category or V-category, is introduced by ~\cite{EiKe66,Lawv73},
where distances became hom-objects. In this paper the definition of
database distance in the V-category $DB$ (which is a strictly
symmetric monoidal category $(DB, \otimes, \Upsilon)$) is
different, as we can see, for example, for every $A\neq\Upsilon$, $d(A,A) = \Upsilon \supset A^A$.\\
\begin{propo} \label{prop:metric} The transitivity law for distance relation
$\sqsubseteq$, and the triangle inequality $d(A,B)\otimes d(B,C)
\sqsupseteq d(A,C)$ for a database metric space are valid. Moreover,
\begin{itemize}
  \item There exists  strong connection between the database
  PO-relation $'\preceq'$ and the distance PO-relation $'\sqsubseteq'$\\
  $~~A \preceq B ~~~~$ iff $~~~~\forall (C\ncong A)(d(A,C) \sqsupseteq
  d(B,C))$, thus\\
  $~~A \simeq B~~~~$ iff $~~~~\forall C(d(A,C) = d(B,C))$
  \item The distances in DB are \textsl{locally closed}. That is, for each object $A$ there exists the bijection\\
  $~~\phi:\{d(A,B)~|~B\ncong A\}~\simeq DB(A,A)$\\
  where $DB(A,A)$ is the hom set of all
  endomorphisms of $A$.
\end{itemize}
\end{propo}
\textbf{Proof:} The transitivity of $\sqsubseteq$ holds because it
is inverse set inclusion relation. Let us show the triangle
inequality:\\
1. case when $A\simeq C$, then $d(B,C) = d(B,A) = d(A,B)$, thus
$d(A,B)\otimes d(B,C) = d(A,B) \sqsupseteq \Upsilon = d(A,C)$.\\
2. case when $A\ncong B$, then\\
2.1 case $A \simeq B$, then $\Upsilon\otimes d(B,C) = d(B,C) =
d(A,C)$(by $A\simeq B$), i.e., $d(A,B)\otimes d(B,C) = d(A,C)$.\\
2.2 case $B \simeq C$, (see 2.1).\\
2.3 case $A\ncong B$ and $B\ncong C$, then $d(A,B)\otimes d(B,C) =
T(A)\bigcap T(B) \bigcap T(C) \subseteq T(A) \bigcap T(C) = d(A,C)$,
i.e., $d(A,B)\otimes d(B,C) \sqsupseteq d(A,C)$.
\\$\square$\\
Notice  that locally closed property means that for any distance
$d(A,B)$ from a database $A$, we have a morphism $\varphi(d(A,B))
= f:A \rightarrow A$, such that $d(A,B) = \widetilde{f}$.\\
From the definition of distance we have that for the infinite
distance $\perp^0$ (which is the terminal and initial object in $DB$
category; denominated infinite object also) we obtain:
$d(\perp^0,\perp^0) = \Upsilon$ (zero distance is the total object
in $DB$ category), and for any other database $A \ncong \perp^0$,
$d(A,\perp^0) = \perp^0$, the distance from $A$ to the infinite
object (database) is \emph{infinite}. Thus, the bottom element
$\perp^0$ and the top element $\Upsilon$
in the database lattice are, for this database metric system, infinite and zero distances (closed objects) respectively.\\
 Let us make a comparison  between this database metric space and the  general metric
space (Frechet axioms):
\begin{center}
\begin{tabular}{|c|c|} \hline
$  $ & $ $\\
  Frechet axioms & $DB$ metric space \\
  $  $ & $ $\\
  \hline
  $  $ & $ $\\
  $~~d(A,B)+ d(B,C)\geq d(A,C)~~$ & $~~d(A,B)\otimes d(B,C) \sqsupseteq d(A,C)~~$
  \\
  $  $ & $ $\\
  $0 \geq d(A,A)$ & $\Upsilon = d(A,A)$ \\
   if $d(A,B) = 0$ then $A = B$ & if $d(A,B) = \Upsilon$ then $A\simeq B$ \\
  $  $ & $ $\\
  $d(A,B) <\infty$ & $d(A,B)\sqsubseteq \bot^0$ \\
  $  $ & $ $\\
  $d(A,B) = d(B,A)$ & $d(A,B) = d(B,A)$ \\ \hline
\end{tabular}
\end{center}
\subsection{Subobject classifier}
Every subset $A \subseteq B$ in the category $Set$ can be described
by its characteristic function $C_f:B \longrightarrow \Omega$, such
that $C_f(x) = True$ if $x\in A$, $False$ otherwise, where $\Omega =
\{True, False\}$ is the set of truth values. In order to generalize
this idea for any two database instances $A \preceq B$ (i.e., a
monomorpfism $f:A \hookrightarrow B$) in $DB$ category, the
subobject classifier $\Omega$ or truth-value object in $DB$ will now
be defined.
\begin{propo}\label{def:sobobject-classifier} Subobject classifier for $DB$ is the object
$\Omega= \Upsilon$ with the arrow $~~true:\perp^0\longrightarrow
\Omega$ that satisfies the
$\Omega$-axiom:\\
For each monomorpfism $in_A:A \hookrightarrow B$ there is one and
only one characteristic arrow $C_{in_A}:B\longrightarrow \Omega$,
where $C_{in_A} = C_{Tin_A}\circ is_B$, with $is_B:B\longrightarrow
TB$ an isomorphism and $C_{Tin_A}:TB \longrightarrow \Omega$  the
characteristic arrow for $Tin_A:TA\hookrightarrow TB$, $~C_{Tin_A}
\triangleq \{id_{\perp}\}\bigcup_{\partial_0(q_{TB_i}) \in TB-TA}
~\{q_{TB_i}\} $, $~~$ such that the diagram
\begin{diagram}
   A  & \rTo^{in_A}   & B \\
 \dTo^{t_A} &       &   \dTo_{C_{in_A}}\\
 \perp^0  &  \rTo^{true}    &    \Omega = \Upsilon \\
 \end{diagram}
is a pullback square.\\
Thus, $DB$ is a monoidal elementary topos.
\end{propo}
\textbf{Proof:} Let us verify that this pullback square commutes.\\
The arrow $t_A:A\twoheadrightarrow \perp^0$ is a unique arrow from
$A$ to the terminal object $\perp^0$, while the arrow
$~~true:\perp^0\longrightarrow \Omega$ (such that $\partial_0(true)
= \partial_1(true) = \perp$ ) is a unique arrow from the initial
object $\perp^0$ to the subobject classifier $\Omega = \Upsilon$,
 thus, $ \widetilde{true\circ t_A} = \widetilde{true} \bigcap
\widetilde{t_A} = \perp \bigcap \perp = \perp$. While,
$C_{in_A}\circ in_A = C_{Tin_A}\circ is_B \circ in_A$. Thus,
$\widetilde{C_{in_A}\circ in_A} =  \widetilde{C_{Tin_A}\circ is_B
\circ in_A} = \widetilde{C_{Tin_A}} \bigcap \widetilde{is_B} \bigcap
\widetilde{in_A}= (TB-TA)\bigcap TB \bigcap TA = (TB-TA) \bigcap TA
= \perp$, and, consequently, diagram commutes. Let us show that it
is a pullback. For any $h:C\longrightarrow B$ and
$t_C:C\twoheadrightarrow \perp^0$, such that $C_{in_A}\circ h =
true\circ t:C$ it must hold that $\widetilde{h} \bigcap (TB-TA) =
\perp$ and $\widetilde{h} \subseteq TB \bigcap TC$,  thus
$\widetilde{h} \subseteq TB$ and, consequently, $\widetilde{h}
\subseteq TA$. But, in that case, there exists $k:C\longrightarrow
A$ such that $\widetilde{k}\subseteq \widetilde{h}$ (in fact,
$\widetilde{k} \subseteq TA \bigcap TC \subseteq TA$) and $ h =
in_A \circ k$.\\
Let us show that $k$ is unique. In fact, for any other
$k_1:C\longrightarrow A$ such that $h = in_A \circ k_1$ we have
$\widetilde{h} = \widetilde{in_A \circ k_1} = \widetilde{in_A}
\bigcap \widetilde{k_1} = TA \bigcap  \widetilde{k_1} =
\widetilde{k_1}$ (because it holds that $\widetilde{k_1} \subseteq
TA \bigcap TC$), so, $\widetilde{k_1} = \widetilde{k}$, i.e., $k_1
=k$.\\
An elementary topos is a Cartesian Closed category with subobject
classifier. The \emph{monoidal elementary topos} is a Monoidal
Closed category, finitely complete and cocomplete, with hom-object
("exponentiation") and a subobject classifier: all properties which
are satisfied by $DB$ category.
\\$\square$
%
\subsection{Weak monoidal topos}

The standard topos is a Cartesian Closed Category with subobject
classifier, that is a finitely complete and cocomplete category
with exponents and subobject classifier.\\
In the previous chapter we defined the database category $DB$ as the
\emph{weak monoidal} topos, which differs from a standard topos by
the fact that, instead of exponents (with cartesian product in the
exponent diagrams),  we have the hom-objects which satisfy the
"exponent" diagrams where the cartesian product $'\times'$ is
replaced by the monoidal tensor product $'\otimes'$.
Let us now compare these two kinds of toposes. \\
In the weak monoidal topos $DB$ the following standard topos
properties that all monomorphisms and epimorphisms are regular: \\
$~~~~~~~~~~~~~~ISOMORPHIC ~~\equiv~~MONIC + EPIC$ \\
$~~~~~~~~~~~~~~EQUALIZER~~\equiv~~MONIC$\\
Recall that in every category is valid  $'ISOMORPHIC ~~imply~~MONIC
+ EPIC'$ and $'EQUALIZER~~imply~~MONIC'$ only.
\begin{propo} \label{prop:monic-equalizer}
  If $f:A\hookrightarrow B$ is a monic arrow then f is an
  equalizer of $C_{in_A}:B\rightarrow\Omega$ and $true_B =
  true\circ t_B:B\rightarrow\Omega$, where $t_B:B \rightarrow
  \perp^0$ is a terminal arrow for B.
\end{propo}
 \textbf{Proof:} Easy to verify.
\\$\square$
\begin{propo} \label{prop:smalest-subobject} $\widetilde{f}$ is the smallest subobject of B
through which $f:A \rightarrow B$ factors. That is, if $f = l\circ
h$ for any $h:A \rightarrow C$ and monic $l:C \hookrightarrow B$,
then there is a unique $k:\widetilde{f}\rightarrow C$ making
\begin{diagram}
   A  & \rOnto^{\tau_f}   & \widetilde{f}
 \\
 \dTo^{h} &  \ldTo^k     &   \dInto_{\tau_f^{-1}}\\
 C  &  \rInto^{l}    &    B \\
\end{diagram}
commute, and hence $\widetilde{\tau_f^{-1}} \subseteq
\widetilde{l}$.
\end{propo}
\textbf{Proof:} $\widetilde{f} = \widetilde{h} \bigcap
\widetilde{l}$, thus $\widetilde{h} \supseteq \widetilde{f}$ and $TC
\supseteq \widetilde{l} \supseteq \widetilde{f}$. From
$\widetilde{f} \supseteq \widetilde{k}$ and $\widetilde{\tau_f^{-1}}
= \widetilde{f} = \widetilde{k} \bigcap \widetilde{l}$, i.e.,
$\widetilde{k} \supseteq \widetilde{f}$, we obtain that
$\widetilde{k} = \widetilde{f}$, thus $k$ is the unique
monomorphisms.
\\$\square$
\begin{propo}\label{prop:coproduct-pullbacks} Coproduct preserve pullbacks. If
\begin{diagram}
   A  & \rTo^f   & D       &   & A_1 & \rTo^{f_1}   & D\\
 \dTo^g &       &   \dTo^k &   & \dTo^{g_1} &       &   \dTo^{k}\\
 B  &  \rTo^h    &    E    &   &  B_1  &  \rTo^{h_1}    &    E\\
 \end{diagram}
 are pullbacks in the $DB$ category, than so is
\begin{diagram}
   A+A_1  & \rTo^{[f,f_1]}   & D       &   \\
 \dTo^{g+g_1} &       &   \dTo^k &   \\
 B+B_1  &  \rTo^{[h,h_1]}   &    E    &   \\
 \end{diagram}
 where $[f,f_1] = ep_D\circ (f + f_1)$, with the epimorphism
 $ep_D:D+D \twoheadrightarrow D$, such that $ep_D = \{ i_r:D \rightarrow D ~|~ \partial_0(i_r) = \partial_1(i_r) = r \in
 D \}$.
\end{propo}
\textbf{Proof:} Easy to verify.
\\$\square$\\
Let us now consider the topos properties which are not satisfied in
$DB$ category.
\begin{propo} \label{prop:topos-no-property} The following topos properies in $DB$ category does
not hold:
\begin{itemize}
  \item Pullbacks does not preserve epics.
  \item $DB$ category has no power objects.
  \item $DB$ category is not well-pointed.
\end{itemize}
\end{propo}
\textbf{Proof:}  1. Let
\begin{diagram}
   D= \widetilde{f} \bigcap \widetilde{g}  & \rInto^{h_A}   & A
 \\
 \dInto^{h_B} &       &   \dOnto_{f}\\
 B  &  \rTo^{g}    &    C \\
 \end{diagram}
 be a pullback square with epimorphism $f:A \rightarrow C$, (i.e., $\widetilde{f} =
 TC$), then $D = \widetilde{g} = \widetilde{h_B}$. Thus, for any
 $B$ such that $TB \supset \widetilde{g} = D$, $\widetilde{h_B}
 \subset TB$, so $h_B$ is not an epimorphism.\\
 2. By definition, the power object of $A$ (if it exists) is an
 object $\P(A)$ which represents the contravariant functor $Sub(\_
 \times A):DB \rightarrow Set$, where for any object $B$, $Sub(B)= \{\widetilde{f} | f$ is a subobject of $A\} =
  \{\widetilde{f} | \widetilde{f} \subseteq TA \}$ is the set of all subobjects
 (monomorphic arrows with the target object $B$) of $B$. Let us show
 that for any object $A \ncong \perp^0$ there is no the power object
 $\P(A)$ such that in $Set$ holds the bijection $DB(\_~, \P(A)) \simeq Sub(\_
 \times A)$. In fact, $Sub(B \times A) = Sub(B + A)= \{\widetilde{f} | \widetilde{f} \subseteq T(A+B)= TA + TB \}
 = Sub(A)+ Sub(B)$. So, $ |DB(B,\P(A))| =|\{\widetilde{f} | \widetilde{f} \subseteq TB\bigcap T(\P(A))\}|
 \subseteq |\{\widetilde{f} | \widetilde{f} \subseteq TB\}|
 \subseteq|Sub(B\times A)|$.\\
 3. The extensionality principle for arrows "if $f,g:A\rightarrow
 B$ is a pair of distinct parallel arrows, then there is an
 element $x:\perp^0 \rightarrow A$ of $A$ such that $f\circ x \neq g
 \circ x$ " does not hold, because $\widetilde{f\circ x} = \widetilde{g \circ
 x} = \perp^0$ for the (unique) element (arrow) $x:\perp^0 \rightarrow
 A$ such that $\widetilde{x} = \perp^0$.
\\$\square$
\section{Conclusions}
In previous work we defined a category $DB$ where objects are
databases and morphisms between them are extensional GLAV mappings
between databases. We defined equivalent (categorically isomorphic)
objects (database instances) from the \emph{behavioral point of view
based on observations}:  each arrow (morphism) is composed by a
number of "queries" (view-maps), and each query may be seen as an
\emph{observation} over some database instance (object of $DB$).
Thus, we  characterized each object in $DB$ (a database instance) by
its behavior according to a given set of observations. In this way
two databases $A$ and $B$ are equivalent (bisimilar) if they have
the same set of its observable internal states, i.e. when $TA$ is
equal to $TB$. It has been shown that such a $DB$ category is equal
to its dual, it is symmetric in the way that the semantics of each
morphism is an closed object (database) and viceversa each database
can be represented by its identity morphism, so that $DB$ is a
2-category.\\
 In this paper we considered some Universal
algebra considerations and relationships of $DB$ category and
standard $Set$ category. We   introduced the categorial (functors)
semantics for two basic database operations: matching and merging
(and data federation), and we defined the algebraic database
lattice.\\
After that we have shown that $DB$ is concrete, small and locally
finitely presentable (lfp) category, and that $DB$ is also monoidal
symmetric V-category enriched over itself. \\
Based on these results  we developed a metric space and a subobject
classifier for this category, and we have shown that it is a weak
monoidal topos.\\
Finally we have shown some negative results for $DB$ category: it is
not well-pointed, it has no power objects, and its pullbacks does
not
preserve epics.\\
  These, and some other, results suggest the need for further investigation
  of categorial coalgebraic semantics for GLAV database mappings
  based on monads, and of general (co)algebraic and (co)induction
  properties for databases.


\bibliographystyle{IEEEbib}

\bibliography{mydb}


\end{document}